\documentclass[aps,prl,reprint,superscriptaddress,amsmath,amssymb,floatfix,longbibliography]{revtex4-2}
\usepackage{graphicx}
\usepackage{bm}
\usepackage{xcolor}
\usepackage[colorlinks=true,linkcolor=blue!60!black,citecolor=blue!60!black,urlcolor=blue!60!black]{hyperref}

\newcommand{\br}[3]{(#1|#2|#3)}
\newcommand{\GR}{\mathrm{GR}}
\newcommand{\sgn}{\operatorname{sgn}}
\newcommand{\Rk}{\mathcal{R}}

\begin{document}

\title{Interferometric trilemma for ultraviolet-softened gravity}

\author{Hiroki Matsui}
\email{hiroki.matsui@yukawa.kyoto-u.ac.jp}
\affiliation{Osaka Central Advanced Mathematical Institute (OCAMI), Osaka Metropolitan University, Osaka 558-8585, Japan}
\affiliation{TRIP Headquarters, RIKEN, Wako, Saitama 351-0198, Japan}

\date{\today}

\begin{abstract}
Ultraviolet-softened extensions of general relativity can weaken gravity at short distances, whereas a positive K\"all\'en--Lehmann  spectral density forbids gravity from falling below Newtonian strength. Which property of the mediator must such a weakening sacrifice? 
For Lorentz-invariant Gaussian mediators coupled linearly to the conserved stress tensor, we formulate three independent interferometric conditions that jointly imply this bound: visibility is never amplified, decoherence is fixed by the energy radiated by the difference source, and no interferometer responds to a later one. At tree level, the three ghost prescriptions of quadratic gravity each violate exactly one: negative-norm ghosts amplify coherence, negative-energy ghosts decohere without positive energy transfer from the difference source, and a time-symmetric fakeon response produces advanced fringe shifts. Any universal short-distance weakening of gravity therefore certifies that its mediator violates one of these conditions, and the observables proposed to certify quantum gravity become consistency tests of its ultraviolet completion.
\end{abstract}

\maketitle

\paragraph{Introduction.}---General relativity (GR) is perturbatively nonrenormalizable, but adding the curvature-squared terms $R^2$ and $C_{\mu\nu\rho\sigma}C^{\mu\nu\rho\sigma}$ makes the theory power-counting renormalizable~\cite{Stelle:1976gc}. These terms change the large-momentum behavior of the graviton propagator from $1/k^2$ to $1/k^4$, thereby modifying the interaction between static masses~\cite{Stelle:1977ry},
\begin{equation}
V_G(r)=-\frac{Gm_Am_B}{r}\Big(1-\tfrac43 e^{-m_2r}+\tfrac13e^{-m_0r}\Big),
\label{eq:Stelle}
\end{equation}
where $m_2$ and $m_0$, set by the coefficients of the $C_{\mu\nu\rho\sigma}C^{\mu\nu\rho\sigma}$ and $R^2$ terms, are the masses of the massive spin-2 and spin-0 modes they introduce. The Newtonian $1/r$ singularity cancels, and $V_G$ stays finite as $r\to0$~\cite{Modesto:2014eta}. This cancellation illustrates how improved ultraviolet (UV) behavior can soften the static potential. In quadratic gravity, this cancellation involves a massive spin-2 mode whose propagator residue is negative, i.e., a ghost. How to quantize the ghost is still debated. It may be given negative norm~\cite{Lee:1969fy,Lee:1970iw,Salvio:2015gsi,Kubo:2023lpz}, negative energy~\cite{Woodard:2015zca,Cline:2003gs}, or no asymptotic states at all, as a purely virtual ``fakeon''~\cite{Anselmi:2018kgz,Anselmi:2018tmf}. Alternatives include an unstable mode with reversed causality~\cite{Donoghue:2019ecz}, complex Lee--Wick pairs~\cite{Cutkosky:1969fq,Grinstein:2008bg}, and nonlocal form factors that soften Eq.~\eqref{eq:Stelle} without new poles~\cite{Biswas:2011ar}. Nonperturbative spectral reconstructions~\cite{Fehre:2021eob} raise the same issue. The debate has largely turned on asymptotic states: whether the ghost appears among them, with what norm or energy, and whether the $S$ matrix is then unitary.

Tabletop proposals to test the quantum nature of gravity use a different language. Gravitationally induced entanglement~\cite{Bose:2017nin,Marletto:2017kzi}, graviton-induced decoherence~\cite{Matsui:2026skx,Matsui:2026abk,Danielson:2021egj,Kanno:2020usf}, and analyses of retardation and the field commutator~\cite{Martin-Martinez:2022uio} all concern, at finite times, the influence functional of the gravitational field on superposed matter. We characterize that functional by the response, commutator, and noise kernels; the commutator kernel is the antisymmetric part of the response. A softened potential is a statement about the conservative part of the response only. We ask whether it constrains the other two, and hence what tabletop observables can reveal about how the ghost is treated. Closed interferometric histories provide a common finite-time framework for comparing prescriptions with different asymptotic state spaces within the Gaussian kernel description below.

For mediators with a positive K\"all\'en--Lehmann spectral density, every massive state contributes an attractive Yukawa term to the Newtonian potential, preventing gravity from becoming weaker than in GR~\cite{Dvali:2010ue}. Equation~\eqref{eq:Stelle} evades this bound because the ghost enters the propagator with negative weight. 
However, a negative residue is not an observable property in itself. A negative-energy ghost has a positive norm and a fakeon has no asymptotic states, yet both reproduce Eq.~\eqref{eq:Stelle}.
The dispute between competing prescriptions is exactly whether a theory with ghosts must give up any physical principle, and if so, which one. 
The K\"all\'en--Lehmann argument implies only that one of its premises fails; for a given classical action, it does not decide which. Our answer is a trilemma. We formulate three conditions on the influence functional: a visibility positivity condition (P), a decoherence--radiation identity (D), and causality (C). Together, under the assumptions stated below, they reproduce the positive spectral form of the static potential, and with it the bound of Ref.~\cite{Dvali:2010ue}. 
The conventional K\"all\'en--Lehmann representation assumes
positive-norm states and a positive-energy spectrum
~\cite{Kallen:1952zz,Lehmann:1954xi,Weinberg:1995mt}.
Our conditions probe the noise, its relation to the energy spectrum, and the relation between the commutator and the static response. Every UV-softened mediator within this class therefore violates at least one condition. The three free ghost prescriptions studied here each violate exactly one of these conditions, with distinct interferometric signatures that we calculate for quadratic gravity.

\paragraph{Influence functional.}---Let branch $a$ of a set of matter interferometers carry a conserved stress tensor $T_a^{\mu\nu}(x)$, with all branches coinciding before splitting and after recombination (a closed history). For a mediator $h_{\mu\nu}$ coupled as $H_{\rm int}=-\tfrac{\kappa}{2}\int d^3x\,h_{\mu\nu}T^{\mu\nu}$ ($\kappa^2=32\pi G$, $\hbar=c=1$), 
we use the Fock vacua for ordinary fields and causal ghosts, and the projected kernels of Sec. S3 for fakeons. The reduced matter density matrix after recombination is~\cite{Feynman:1963fq,Matsui:2026skx}
\begin{align}
\rho_{ab}=\rho^{(0)}_{ab}\exp\Big\{&\frac{\kappa^2}{4}\int d^4x\,d^4x'\,\Delta_{ab}(x)\Big[i\Rk(x,x')\,\Sigma_{ab}(x')\nonumber\\
&-\tfrac12G_{\rm H}(x,x')\,\Delta_{ab}(x')\Big]\Big\},
\label{eq:IF}
\end{align}
where $\Delta_{ab}=T_a-T_b$ and $\Sigma_{ab}=\tfrac12(T_a+T_b)$. Tensor indices are suppressed here and below; they are contracted as $\Delta_{ab}^{\mu\nu}(x)\Rk_{\mu\nu,\rho\sigma}(x,x')\Sigma_{ab}^{\rho\sigma}(x')$. 
Here $\Delta$ and $\Sigma$ are the difference and average of the two branch sources. The first term in Eq.~\eqref{eq:IF}  determines their relative phase through the field response, while the second controls the interference visibility through field fluctuations. This Gaussian form follows from trace preservation and Hermiticity (Supplemental Material, Sec.~S1). It contains a real response kernel $\Rk$ and a symmetric noise (Hadamard) kernel $G_{\rm H}$. In linearized quantum gravity, $\Rk=G_R\equiv i\theta(t-t')\langle[h(x),h(x')]\rangle$ and $G_{\rm H}=\tfrac12\langle\{h(x),h(x')\}\rangle$, and Eq.~\eqref{eq:IF} is exact for a free field in a Gaussian state with linear coupling. For two branches it reduces to the coherent-displacement result of Ref.~\cite{Matsui:2026skx}. We define the commutator (Pauli--Jordan) kernel $G_{\rm PJ}\equiv \Rk-\Rk^{T}$, with $\Rk^{T}(x,x')\equiv\Rk(x',x)$, which equals $i\langle[h,h]\rangle$ for the operator realization considered here. 
For fakeons, we use Eq.~\eqref{eq:IF} with the projected kernels of Sec.~S3, assuming that this prescription also applies to finite-time interferometry.

Two observables follow (Sec.~S1). The first is the visibility loss of one interferometer, $|\rho_{12}|=|\rho^{(0)}_{12}|\,e^{-\Gamma}$ with
\begin{equation}
\Gamma=\frac{\kappa^2}{8}\int d^4x\,d^4x'\,\Delta(x)\,G_{\rm H}(x,x')\,\Delta(x').
\label{eq:Gamma}
\end{equation}
The second observable is the change in the interference phase of $B$ when $A$ switches between its two branches. Denoting the stress-tensor differences of $A$ and $B$ by $X$ and $Y$, respectively, Eq.~\eqref{eq:IF} gives
\begin{equation}
\delta\phi_{B|A}=\frac{\kappa^2}{4}\int d^4x\,d^4x'\,Y(x)\,\Rk(x,x')\,X(x').
\label{eq:cond}
\end{equation}
The shift $\delta\phi_{A|B}$ follows by exchanging $X$ and $Y$. Here $\delta\phi_{B|A}$ is the shift that an experimenter operating $B$ observes in the fringe pattern when $A$ is switched from branch $0$ to branch $1$. Measuring it requires no entanglement, since $A$ can be prepared in a definite branch. It is the interferometric version of a response function, with $\Rk(x,x')$ giving the field at $x$ on $B$'s history produced by $A$'s source at $x'$. For interferometers operated at different times, comparing $\delta\phi_{B|A}$ with $\delta\phi_{A|B}$ reveals the time ordering of that response.

We consider mediators whose noise and commutator kernels can be decomposed into contributions from species of different masses and spins,
\begin{equation}
G_{\rm H}=\sum_s\nu_s G_{{\rm H},s},\qquad G_{\rm PJ}=\sum_s\varepsilon_s G_{{\rm PJ},s}.
\label{eq:spectral}
\end{equation}
Here $s$ labels a spin-0 or spin-2 species of mass $\mu_s$, and $G_{{\rm H},s}$ and $G_{{\rm PJ},s}$ are the kernels of an ordinary particle with positive norm and positive energy. The constants $\nu_s$ and $\varepsilon_s$ are the noise and energy-transfer weights of these vacuum reference kernels. We normalize the ordinary massless graviton to $\nu_g=\varepsilon_g=1$; an ordinary massive species likewise has unit weights. For a causal spin-2 ghost, however, $\varepsilon_2=-1$, while $\nu_2$ depends on how the ghost is quantized. The response kernel $\Rk$ can in addition depend on the propagation prescription, as for a fakeon.

\paragraph{The trilemma.}---We impose three conditions on Eq.~\eqref{eq:IF} and translate each into the spectral data (proofs in Sec.~S2). Under Eq.~\eqref{eq:spectral}, decoherence and the difference-source radiation functional take a simple form. Let $n_s[\Delta]\geq0$ be the mean number of quanta that an ordinary species $s$ would emit from the difference source $\Delta$, and let $\Gamma_{\rm rad}\equiv\int d\omega\,(2\omega)^{-1}dW[\Delta]/d\omega$, where $W[\Delta]$ is the net energy transferred to the mediator by the difference source (Secs.~S1 and S2). Both $n_s[\Delta]$ and $W[\Delta]$ are quadratic functionals of $T_a-T_b$, rather than the radiation of either branch; they contain interference terms between the branch sources. Then
\begin{equation}
\Gamma=\tfrac12\sum_s\nu_s\,n_s[\Delta],\qquad
\Gamma_{\rm rad}=\tfrac12\sum_s\varepsilon_s\,n_s[\Delta],
\label{eq:GW}
\end{equation}
so the noise weights $\nu_s$ govern decoherence and the commutator weights $\varepsilon_s$ govern radiation. The conditions are imposed on all conserved closed histories of an admissible source class (see the assumptions below).

(P) \emph{Visibility positivity}: no closed history increases the visibility of a two-path interferometer, $\Gamma\geq0$. This is necessary for positivity of the reduced matter state, but is not sufficient for positivity of an arbitrary multibranch Gaussian map. By Eq.~\eqref{eq:GW}, $\mathrm{P}\Leftrightarrow\nu_s\geq0$.

(D) \emph{Decoherence--radiation identity}: $\Gamma=\Gamma_{\rm rad}$ for every closed history. In GR this is $\Gamma=N_\Delta/2$, with $N_\Delta=n_g[\Delta]$ the mean graviton occupation associated with the difference source~\cite{Matsui:2026skx}. D is the zero-temperature fluctuation--dissipation relation for the reference kernels. In a positive-metric operator realization it expresses the positive-frequency spectral condition in the source-coupled sector; it does not by itself establish stability of the full interacting theory. By Eq.~\eqref{eq:GW}, $\mathrm{D}\Leftrightarrow\nu_s=\varepsilon_s$.

(C) \emph{Causality}: an interferometer's fringe is unaffected by an interferometer operated entirely in its future, $\delta\phi_{B|A}=0$ if $A$ is later than $B$. No operation performed later can then be detected earlier, which is the operational meaning of retarded propagation. For a sufficiently rich class of conserved probes, Eq.~\eqref{eq:cond} implies that the observable response vanishes for $t<t'$. Away from coincident points, this fixes $\Rk=\theta G_{\rm PJ}$. This equality is understood on conserved sources, modulo local contact terms. Such terms do not affect the potential between separated point sources ($r>0$).

\emph{Theorem.} For a well-defined Lorentz-invariant response with only the ordinary graviton in its massless sector and the spectral representation~\eqref{eq:spectral}, P, D, and C imply, on a sufficiently rich class of conserved probes, that the static potential at $r>0$ takes the positive spectral (K\"all\'en--Lehmann) form
\begin{align}
V_G(r)&=-\frac{G_\infty m_Am_B}{r}\Big[1+\sum_{\mu_s>0}\sigma_s\,e^{-\mu_s r}\Big],\nonumber\\
\sigma_s&=\frac{\varepsilon_s\, g_s}{G_\infty}=\frac{\nu_s\, g_s}{G_\infty}\geq0,
\label{eq:theorem}
\end{align}
with the sum replaced by an integral for a continuum. Here $g_s>0$ is the static coupling of species $s$, defined by the Yukawa potential $-g_sm_Am_Be^{-\mu_sr}/r$ that an ordinary species of mass $\mu_s$ would mediate on its own (Sec.~S2; in quadratic gravity $g_g=G$, $g_2=\frac43G$, and $g_0=\frac13G$, cf.\ Eq.~\eqref{eq:Stelle}), and $G_\infty=\lim_{r\to\infty}r|V_G|/(m_Am_B)$ is Newton's constant measured at long distances, set by the ordinary massless graviton sector assumed here. Hence both the potential and the force are at least Newtonian at every distance,
\begin{equation}
|V_G(r)|\geq \frac{G_\infty m_Am_B}{r},\qquad |V_G'(r)|\geq \frac{G_\infty m_Am_B}{r^2}.
\label{eq:bound}
\end{equation}

\emph{Proof.} For static point masses held for a time $T$, the potential is defined by $\delta\phi_{B|A}=-V_G(r)\,T$, and $\Rk$ enters only through $\int_{-\infty}^{\infty}d\tau\,\Rk(\tau,r)$. Under C, this equals $\int_0^\infty d\tau\,G_{\rm PJ}(\tau,r)$, so $V_G(r)=-\frac{\kappa^2}{4}m_Am_B\int_0^\infty d\tau\,(G_{\rm PJ})_{00,00}(\tau,r)$. Inserting Eq.~\eqref{eq:spectral}, each species $s$ contributes $\varepsilon_s$ times the Yukawa term $-g_sm_Am_Be^{-\mu_sr}/r$ of an ordinary species, where $g_s>0$ because static exchange of even-spin quanta between positive masses is attractive (Sec.~S2). This gives Eq.~\eqref{eq:theorem}, with the massless sector supplying the Newtonian term, and P and D give $\varepsilon_s=\nu_s\geq0$. Since $e^{-\mu r}$ and $(1+\mu r)e^{-\mu r}$ are positive, Eq.~\eqref{eq:bound} follows term by term, which completes the proof.

C fixes $V_G$ by the $\varepsilon_s$, D sets $\varepsilon_s=\nu_s$, and P gives $\nu_s\geq0$, hence $\sigma_s\geq0$. Every exchange that survives is attractive, so gravity can only grow stronger at shorter distances. Equivalently, $k^2|\tilde V_G(k)|$ cannot decrease, so a $1/k^4$ falloff of the exchange part is impossible. The bound~\eqref{eq:bound} itself is not new. It is the K\"all\'en--Lehmann obstruction~\cite{Kallen:1952zz,Lehmann:1954xi,Weinberg:1995mt}, which Ref.~\cite{Dvali:2010ue} applied to gravity, showing that a ghost-free, weakly coupled theory cannot be weaker than GR. Our contribution is an operational decomposition into P, D, and C: distinct conditions on interferometric kernels that jointly imply the bound and can fail separately. For theories with ghosts, the K\"all\'en--Lehmann argument says only that its premise fails; this decomposition identifies which kernel condition is lost.

The three conditions are independent. Table~\ref{tab:classes} lists ghost prescriptions of quadratic gravity that each violate exactly one, so none of the conditions can be dropped. The \emph{contrapositive} is the main point. Within the assumptions stated at the end of this Letter, any universal (composition-independent) weakening of gravity at short distances certifies that the mediator violates P, D, or C, and the interferometric fingerprints below identify which. This applies directly to Eq.~\eqref{eq:Stelle} in the free prescriptions considered here.

\paragraph{Quadratic gravity.}---The propagator of quadratic gravity contains three species with classical residues $(+,-,+)$: the graviton, a spin-2 mode of mass $m_2$, and a spin-0 mode of mass $m_0$. The static response is fixed by the classical action, so Eq.~\eqref{eq:Stelle} holds in every treatment; for a causal ghost, $\varepsilon=(1,-1,1)$. The prescriptions differ in how the ghost enters the fluctuations and in whether the response is retarded.

A free \emph{negative-norm} ghost has $\nu_2=\varepsilon_2=-1$ and breaks P. For a negative-norm mode the characteristic function of a coherent displacement $\beta$ is $e^{+|\beta|^2/2}$ instead of $e^{-|\beta|^2/2}$, so emitting such a quantum \emph{increases} the overlap of the two branch states. A \emph{negative-energy} ghost has positive norm, so $\nu_2=+1$ and $\varepsilon_2=-1$, and it breaks D. It contributes ordinary positive noise but an oppositely signed commutator and response. In the free, tree-level projected prescription, a \emph{fakeon} has no on-shell contribution, $\nu_2=\varepsilon_2=0$. It neither fluctuates nor radiates, and modifies only the conservative interaction. Its contribution to $\Rk$ is time-symmetric, $-G_{\rm PV}^{(m_2)}=-\frac12(G_R+G_A)^{(m_2)}$ with $G_A$ the advanced Green function, which breaks C.

Related causal behavior has been discussed for reversed-causality ghosts, complex Lee--Wick pairs, and nonlocal theories~\cite{Donoghue:2019ecz,Cutkosky:1969fq,Grinstein:2008bg,Biswas:2011ar,Carone:2016eyp}. Applying our theorem to these models or to spectral reconstructions requires checking Eq.~\eqref{eq:spectral} and the existence of a finite-time response for each prescription (Sec.~S3).

For a slowly moving closed history with difference second mass moment $\Delta I_{ij}(t)$, Eq.~\eqref{eq:Gamma} evaluates to (Sec.~S3; $\hbar$ and $c$ restored, $\mu_s=m_sc^2/\hbar$)
\begin{equation}
\Gamma=\frac{G}{4\pi\hbar c^5}\sum_s\nu_s\int_{\mu_s}^\infty\!d\omega\,\omega^5\,v_s\,\mathcal W_s(v_s),
\label{eq:Gamma-massive}
\end{equation}
with $v_s=(1-\mu_s^2/\omega^2)^{1/2}$ the speed of the emitted quantum in units of $c$ and $\mathcal W_s(v)$ the angular-averaged polarization sum of species $s$, a quadratic form in the Fourier transform $\Delta\tilde I_{ij}(\omega)$ (explicit forms and derivation in Sec.~S3). For the graviton alone, $\mathcal W_g=\frac25|\Delta\tilde I^{\rm TF}|^2$, with TF the trace-free part, and Eq.~\eqref{eq:Gamma-massive} is the quadrupole formula of Ref.~\cite{Matsui:2026skx}. $\Gamma_{\rm rad}$ follows by replacing $\nu_s$ with $\varepsilon_s$. At high frequency the polarization sums obey a \emph{decoherence sum rule},
\begin{equation}
\mathcal W_g-\mathcal W_2(1)+\mathcal W_0(1)=0,
\label{eq:sumrule}
\end{equation}
which is the dynamical counterpart of $1-\frac43+\frac13=0$ in Eq.~\eqref{eq:Stelle}. The leading high-frequency contribution to the $\varepsilon$-weighted radiation functional thus cancels; the noise has the same cancellation when $\nu=\varepsilon$.

\begin{figure}[tb]
\includegraphics[width=\columnwidth]{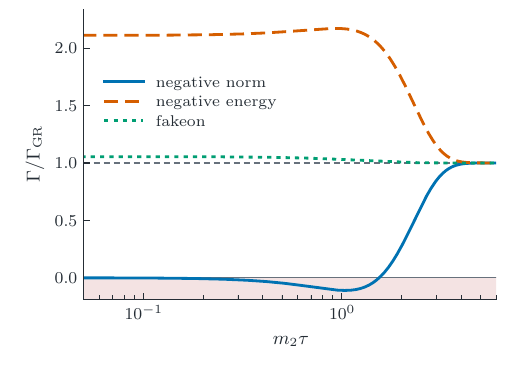}
\caption{Decoherence of the orthogonal-arm history relative to GR versus $m_2\tau$ for the three ghost prescriptions of quadratic gravity ($m_0=m_2$). For both causal ghosts, $\Gamma_{\rm rad}$ coincides with the negative-norm curve; for the fakeon, $\Gamma_{\rm rad}=\Gamma$. Shaded: $\Gamma<0$ (coherence amplification).}
\label{fig:gamma}
\end{figure}

\begin{figure}[tb]
\includegraphics[width=\columnwidth]{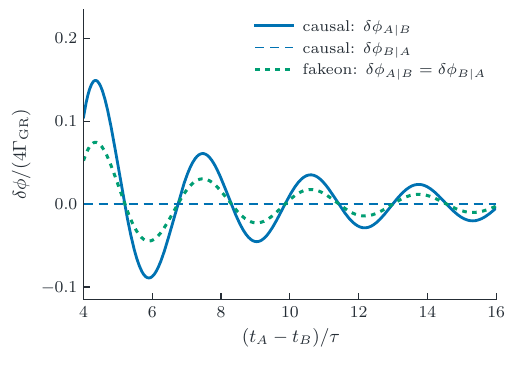}
\caption{Ghost wake component inferred from the commutator bilinear for two orthogonal-arm profiles with delay $t_A-t_B\geq4\tau$ ($m_2\tau=2$, $m_0=m_2$). The curves use the separated-support relations~\eqref{eq:reciprocal}: the causal wake acts forward, and the fakeon wake is shared equally. Gaussian tails overlap, so the curves omit overlap-dependent conservative response terms and are not the full conditional shifts. The omitted massless commutator bilinear is below $0.025$ in these units (Sec.~S5).}
\label{fig:wake}
\end{figure}

We evaluate Eq.~\eqref{eq:Gamma-massive} for an idealized orthogonal-arm history, in which the two branches undergo Gaussian excursions $\bm r_{1,2}=d\,e^{-t^2/2\tau^2}\hat{\bm x},\,d\,e^{-t^2/2\tau^2}\hat{\bm y}$, assuming that the total conserved source, including the apparatus, has the stated particle-dominated second-moment difference. Then $\Delta I_{ij}$ is trace-free and $\Gamma_\GR=\frac85Gm^2d^4/\hbar c^5\tau^4$. The scalar couples mainly to the trace of $\Delta I_{ij}$, so its high-frequency weight relative to the graviton is small, $\mathcal W_0(1)/\mathcal W_g=\frac1{18}$, while $\mathcal W_2(1)/\mathcal W_g=\frac{19}{18}$ (the collinear history of Ref.~\cite{Matsui:2026skx} is treated in Secs.~S3--S5). When both $m_2\tau\gg1$ and $m_0\tau\gg1$, massive emission is confined to the spectral tails and all prescriptions approach GR. The spin-2 correction scales as $(m_2\tau)^3e^{-(m_2\tau)^2/2}$; the scalar correction is controlled separately by $m_0\tau$. For $m_2\tau\lesssim2$ they separate (Figs.~\ref{fig:gamma} and \ref{fig:wake}, Table~\ref{tab:classes}).

\emph{P broken (negative norm).} Because $\nu=\varepsilon$, the noise obeys the sum rule~\eqref{eq:sumrule}, and the ghost overcompensates in a window above $m_2$: for $m_0=m_2$ we find $\Gamma<0$ whenever $m_2\tau<1.55$, with a minimum of $\Gamma/\Gamma_\GR=-0.11$, and for every $m_0/m_2\in[0.1,\infty)$ that we examined the threshold lies between $m_2\tau=1.45$ and $1.57$ (Sec.~S4). Such amplification violates the unit visibility bound for an initially fully coherent balanced input.

\emph{D broken (negative energy).} The noise adds instead of canceling: $\Gamma/\Gamma_\GR\to19/9$ as $m_2\tau\to0$. Meanwhile the $\varepsilon$-weighted quantities cancel by Eq.~\eqref{eq:sumrule}: $\Gamma_{\rm rad}$ equals the negative-norm value of $\Gamma$ (Fig.~\ref{fig:gamma}), and the difference-source radiative work $W[\Delta]$ vanishes or even turns negative (Sec.~S4). The mismatch $\Gamma-\Gamma_{\rm rad}$ is exactly the difference-source ghost occupation $n_2[\Delta]$ (Sec.~S4). This is the structure of a quantum-mechanics-free subsystem built from positive- and negative-mass oscillators~\cite{Tsang:2012nt,Moller:2017yzv}, in which the commutators, and with them the back-action, cancel while the noise adds.

\emph{C broken (fakeon).} Here $\nu$ and $\varepsilon$ contain only healthy species, so D holds and $\Gamma_{\rm rad}=\Gamma$. Decoherence is then that of GR plus the scalar, $\Gamma/\Gamma_\GR\to19/18$. The ghost survives only in the time-symmetric part of $\Rk$; its advanced response can be isolated using conditional shifts of temporally ordered, timelike-separated histories. For spatially compact histories with timelike-separated supports and $A$ entirely later than $B$, massless GR gives no interaction at all (Huygens' principle), so any coupling comes from the massive wake inside the light cone. For histories with disjoint temporal support, the ghost contributions to the two shifts are (Sec.~S5)
\begin{equation}
\delta\phi^{\rm causal}_{B|A}=0,\qquad
\delta\phi^{\rm fakeon}_{B|A}=\delta\phi^{\rm fakeon}_{A|B}=\tfrac12\,\delta\phi^{\rm causal}_{A|B},
\label{eq:reciprocal}
\end{equation}
where ``causal'' refers to the negative-norm and negative-energy ghosts, whose wake acts only forward in time. For a fakeon, the earlier interferometer's fringe responds to the later choice of branch, with the massive wake shared equally between past and future, as in Wheeler--Feynman exchange~\cite{Wheeler:1945ps,Anselmi:2020tqo}. When the scalar wake is negligible and the denominator is nonzero, the ratio $\delta\phi_{B|A}/\delta\phi_{A|B}$ of the full shifts is $0$ for a causal mediator and approximately $1$ for a fakeon (Sec.~S5). The scalar correction can matter near nodes of the spin-2 wake. The conditional shifts thus provide an in-principle diagnostic of advanced response in this projected prescription~\cite{Anselmi:2018tmf}.

\begin{table}[tb]
\caption{Classification of quadratic gravity by the broken condition, for the orthogonal-arm history with $m_0=m_2$. $\Gamma/\Gamma_\GR$ and $\Gamma_{\rm rad}/\Gamma_\GR$ are the limits $m_2\tau\to0$; $0^-$ means that zero is approached from below. Condition D requires $\Gamma=\Gamma_{\rm rad}$. The last column refers to timelike-separated histories with $A$ later than $B$, neglecting the scalar wake and assuming a nonzero denominator (Sec.~S5). The GR ratio is undefined because both shifts vanish by Huygens' principle.}
\label{tab:classes}
\begin{ruledtabular}
\begin{tabular}{lcccc}
ghost & broken & $\Gamma/\Gamma_\GR$ & $\Gamma_{\rm rad}/\Gamma_\GR$ & $\delta\phi_{B|A}/\delta\phi_{A|B}$\\
\hline
none (GR) & --- & $1$ & $1$ & ---\\
negative norm & P & $0^-$\,\footnote{$\Gamma<0$ for $m_2\tau<1.55$.} & $0^-$ & $0$\\
negative energy & D & $19/9$ & $0^-$ & $0$\\
fakeon & C & $19/18$ & $19/18$ & $\simeq1$\\
\end{tabular}
\end{ruledtabular}
\end{table}

\paragraph{Discussion.}---The dynamical fingerprints require $m_2\tau\lesssim2$. The single-Yukawa limit of Ref.~\cite{Lee:2020zjt}, $\lambda<38.6\,\mu$m at gravitational strength, gives an indicative mass scale of $m_2\sim5$\,meV. For the center-of-mass motion of any foreseeable interferometer, $\tau\gg0.2$\,ps, so the deviations in Figs.~\ref{fig:gamma} and \ref{fig:wake} are suppressed by $e^{-(m_2\tau)^2/2}$. The overall scale is also tiny, $\Gamma_\GR\sim10^{-61}$ for parameters of proposed gravitational-entanglement experiments~\cite{Matsui:2026skx}. The trilemma is therefore a structural result and not a near-term signal.

Its potential experimental application lies in the static sector. Short-range force measurements and gravitational-entanglement experiments operating at separations of order $1/m_2$~\cite{Bose:2017nin,vanManen:2026akg} test the bound~\eqref{eq:bound} directly. Within the theorem's assumptions, a universal weakening found there would violate P, D, or C. 

The assumptions are linear (tree-level) coupling to a conserved, universal $T^{\mu\nu}$, Lorentz invariance, the prescribed Gaussian Fock vacuum or projected reference kernels, the ordinary graviton as the only massless sector, noise and commutator kernels built from ordinary mass shells, Eq.~\eqref{eq:spectral}, and an admissible class of conserved histories rich enough to resolve each observable spectral sector. The test tensors in Sec.~S2 establish this richness mathematically; physical realizability is an additional assumption. 
The theorem is not a no-go statement for every modification of gravity: Lorentz-violating completions such as Ho\v{r}ava--Lifshitz gravity~\cite{Horava:2009uw}, which soften the propagator with higher spatial derivatives and no ghosts, are outside its scope.
Momentum-dependent matter vertices change the assumed local coupling, while nonlinear screening takes the response outside the linear regime; neither is covered (Sec.~S7). The required Lorentzian kernel prescription for nonlocal mediator form factors is discussed in Sec.~S3. Loop corrections enter at $O(G^2)$. Within these assumptions, the observables proposed to certify the \emph{quantum} nature of gravity, namely decoherence and the conditional fringe shifts that underlie gravitationally induced entanglement, acquire a second role: they test necessary consistency conditions on the noise, vacuum spectrum, and causal response of its UV completion.

\paragraph{Data availability.}---The numerical results can be reproduced using the formulas provided here and in the Supplementary Material; data is available from the author upon reasonable request.

\bibliography{refs}
\bibliographystyle{utphys}

\clearpage
\onecolumngrid
\setcounter{equation}{0}
\setcounter{table}{0}
\setcounter{figure}{0}
\setcounter{section}{0}
\setcounter{secnumdepth}{2}
\renewcommand{\theequation}{S\arabic{equation}}
\renewcommand{\theHequation}{S\arabic{equation}}
\renewcommand{\thetable}{S\arabic{table}}
\renewcommand{\theHtable}{S\arabic{table}}
\renewcommand{\thefigure}{S\arabic{figure}}
\renewcommand{\theHfigure}{S\arabic{figure}}
\renewcommand{\thesection}{S\arabic{section}}
\renewcommand{\theHsection}{S\arabic{section}}

\begin{center}
{\large\bfseries Supplemental Material}
\end{center}
\medskip

\noindent This Supplemental Material presents the derivations underlying the main-text results. Section~\ref{sec:S1} derives the influence functional and the interferometric observables. Section~\ref{sec:S2} introduces the spectral representation and proves the theorem. Section~\ref{sec:S3} applies the framework to quadratic gravity, Sec.~\ref{sec:S4} presents the numerical results and discusses the radiated energy, and Sec.~\ref{sec:S5} analyzes the reciprocal wake of a fakeon. Section~\ref{sec:S6} discusses the physical scales, and Sec.~\ref{sec:S7} delineates the scope of the assumptions. Equations without the prefix ``S'' refer to the main text. We use $\hbar=c=1$ except in numerical estimates, the mostly-plus metric $\eta_{\mu\nu}={\rm diag}(-1,1,1,1)$, and the Fourier convention $\tilde f(\omega,\bm k)=\int d^4x\,e^{i\omega t-i\bm k\cdot\bm x}f(t,\bm x)$ of Ref.~\cite{Matsui:2026skx}.

\section{Influence functional and interferometric observables}
\label{sec:S1}

\subsection{Setup}
Each interferometer is idealized, as in Ref.~\cite{Matsui:2026skx}, by a set of narrow, approximately orthogonal semiclassical branches $|a\rangle$. Branch $a$ carries a classical, conserved stress tensor $T_a^{\mu\nu}(x)$, $\partial_\mu T_a^{\mu\nu}=0$. The difference $T_a-T_b$ between any two branches is compactly supported in time: the branches coincide before the splitting and after the recombination. We call such a configuration a closed history. The stress tensor includes the apparatus and any branch-dependent recoil or control stresses needed for conservation. Apparatus components common to all branches cancel from $\Delta$ and hence from the noise functional, although they can enter absolute phases through $\Sigma$. Initially the matter and the mediator are uncorrelated,
\begin{equation}
\rho_{\rm in}=\sum_{a,b}\rho^{(0)}_{ab}\,|a\rangle\langle b|\otimes\rho_h ,
\end{equation}
where $\rho_h$ denotes the stationary, zero-mean Gaussian vacuum of the chosen prescription: the stipulated Fock vacuum where an operator realization exists, without assuming that it is a ground state. Projected modes are specified by the kernel ansatz below. The coefficients $\rho^{(0)}_{ab}$ already contain all branch-dependent phases that do not involve the mediator (free evolution, external fields). The mediator couples linearly,
\begin{equation}
H_{\rm int}(t)=-\frac{\kappa}{2}\int d^3x\,h_{\mu\nu}(t,\bm x)T^{\mu\nu}(t,\bm x)\equiv-\int d^3x\,h_{\mu\nu}J^{\mu\nu},\qquad J^{\mu\nu}\equiv\frac{\kappa}{2}T^{\mu\nu},
\label{eq:S-Hint}
\end{equation}
and we write $A_a(t)=\int d^3x\,h_{\mu\nu}J_a^{\mu\nu}$ and $h[J]=\int d^4x\,h_{\mu\nu}J^{\mu\nu}$. We express all bilinears using the bracket notation
\begin{equation}
\br{T}{K}{T'}=\frac{\kappa^2}{4}\int d^4x\,d^4x'\,T^{\mu\nu}(x)K_{\mu\nu,\rho\sigma}(x,x')T'^{\rho\sigma}(x')=\int d^4x\,d^4x'\,J(x)K(x,x')J'(x'),
\end{equation}
where tensor indices are suppressed in the last expression. In this notation, Eqs.~\eqref{eq:IF}, \eqref{eq:Gamma}, and \eqref{eq:cond} read $\rho_{ab}=\rho^{(0)}_{ab}\exp[i\br{\Delta_{ab}}{\Rk}{\Sigma_{ab}}-\frac12\br{\Delta_{ab}}{G_{\rm H}}{\Delta_{ab}}]$, $\Gamma=\frac12\br{\Delta}{G_{\rm H}}{\Delta}$, and $\delta\phi_{B|A}=\br{Y}{\Rk}{X}$. Since branch $a$ evolves with $H_{\rm int}$ evaluated on $T_a$, the joint state after recombination is $\sum_{ab}\rho^{(0)}_{ab}|a\rangle\langle b|\otimes U_a\rho_hU_b^\dagger$, where $U_a$ is the interaction-picture evolution operator of the mediator in branch $a$. Tracing out the mediator gives
\begin{equation}
\rho_{ab}=\rho^{(0)}_{ab}\,{\rm Tr}\big[\rho_h\,U_b^\dagger U_a\big]\equiv\rho^{(0)}_{ab}\,\langle U_b^\dagger U_a\rangle .
\label{eq:S-rhoab}
\end{equation}

\subsection{Operator derivation for a linear field}
Suppose first that $h$ is a free (or generalized free) field, so that its commutator is a c-number. We write it as
\begin{equation}
[h(x),h(x')]=-iG_{\rm PJ}(x,x'),\qquad G_{\rm PJ}(x,x')=i\langle[h(x),h(x')]\rangle ,
\end{equation}
where $G_{\rm PJ}$ is real and antisymmetric. The retarded and advanced Green functions are $G_R(x,x')=\theta(t-t')G_{\rm PJ}(x,x')$ and $G_A(x,x')=G_R(x',x)$, so that $G_{\rm PJ}=G_R-G_A$. We also write $G=\frac12(G_R+G_A)$ for the time-symmetric part.

\emph{Magnus expansion.} With $H_{\rm int}=-A_a(t)$, the evolution operator is $U_a=\mathcal T\exp[i\int dt\,A_a(t)]$. The Magnus expansion gives $U_a=\exp(\Omega_1+\Omega_2+\cdots)$, with
\begin{equation}
\Omega_1=i\!\int\!dt\,A_a(t)=i\,h[J_a],\qquad
\Omega_2=\frac12\int\!dt\!\int^{t}\!dt'\,[iA_a(t),iA_a(t')]=-\frac12\int\!\!\int_{t>t'}dt\,dt'\,[A_a(t),A_a(t')] .
\end{equation}
Every higher term $\Omega_{n\geq3}$ contains nested commutators $[A,[A,A]]$, which vanish because $[A,A]$ is a c-number. Hence the expansion terminates. Using $[A_a(t),A_a(t')]=-i\int d^3x\,d^3x'\,J_a(x)G_{\rm PJ}(x,x')J_a(x')$ we find
\begin{equation}
U_a=e^{ih[J_a]}\,\exp\Big[\frac i2\int\!\!\int d^4x\,d^4x'\,\theta(t-t')J_a(x)G_{\rm PJ}(x,x')J_a(x')\Big]=e^{ih[J_a]}\,e^{\frac i2\br{T_a}{G_R}{T_a}} .
\label{eq:S-U}
\end{equation}
The second factor is a c-number phase; since only the symmetric part of a kernel survives a contraction with the same source, it equals $e^{\frac i2\br{T_a}{G}{T_a}}$.

\emph{Overlap of branch states.} For operators whose commutator is a c-number, $e^Pe^Q=e^{P+Q}e^{\frac12[P,Q]}$. Taking $P=-ih[J_b]$ and $Q=ih[J_a]$ gives $[P,Q]=[h[J_b],h[J_a]]=-i\br{T_b}{G_{\rm PJ}}{T_a}$, so
\begin{equation}
U_b^\dagger U_a=e^{\frac i2\br{T_a}{G}{T_a}-\frac i2\br{T_b}{G}{T_b}-\frac i2\br{T_b}{G_{\rm PJ}}{T_a}}\;e^{ih[J_a-J_b]} .
\end{equation}
We now substitute $T_a=\Sigma+\frac12\Delta$ and $T_b=\Sigma-\frac12\Delta$. Because $G$ is symmetric and $G_{\rm PJ}$ antisymmetric,
\begin{equation}
\br{T_a}{G}{T_a}-\br{T_b}{G}{T_b}=2\br{\Delta}{G}{\Sigma},\qquad
\br{T_b}{G_{\rm PJ}}{T_a}=\tfrac12\br{\Sigma}{G_{\rm PJ}}{\Delta}-\tfrac12\br{\Delta}{G_{\rm PJ}}{\Sigma}=-\br{\Delta}{G_{\rm PJ}}{\Sigma}.
\end{equation}
The total phase is therefore $\br{\Delta}{G}{\Sigma}+\frac12\br{\Delta}{G_{\rm PJ}}{\Sigma}=\br{\Delta}{G_R}{\Sigma}$, and
\begin{equation}
\langle U_b^\dagger U_a\rangle=e^{\,i\br{\Delta}{G_R}{\Sigma}}\;\chi[\Delta],\qquad \chi[\Delta]\equiv\big\langle e^{ih[\kappa\Delta/2]}\big\rangle .
\label{eq:S-overlap}
\end{equation}

\emph{Characteristic functional.} For a Gaussian (quasi-free) state with $\langle h\rangle=0$, all connected correlators beyond the second vanish, and the characteristic functional of the linear operator $\Phi=h[J]$ is $\langle e^{i\Phi}\rangle=e^{-\frac12\langle\Phi^2\rangle}$. Since $\langle\Phi^2\rangle=\frac12\langle\{\Phi,\Phi\}\rangle$,
\begin{equation}
\chi[\Delta]=\exp\Big[-\tfrac12\br{\Delta}{G_{\rm H}}{\Delta}\Big],\qquad G_{\rm H}(x,x')=\tfrac12\langle\{h(x),h(x')\}\rangle .
\end{equation}
Inserting Eq.~\eqref{eq:S-overlap} into Eq.~\eqref{eq:S-rhoab} reproduces Eq.~\eqref{eq:IF} with $\Rk=G_R$. This derivation does not require weak coupling; the result is therefore exact for a linear field. For two branches, it reduces to the displacement-operator formula of Ref.~\cite{Matsui:2026skx}, in which $\chi[\Delta]$ is the characteristic function of the initial graviton state evaluated at the displacement difference.

\subsection{Mode representation}
The same structure can be exhibited mode by mode. For a single canonical oscillator of frequency $\omega$ with field $\phi(t)=(ae^{-i\omega t}+a^\dagger e^{i\omega t})/\sqrt{2\omega}$, the source $J(t)$ enters through
\begin{equation}
h[J]=\int dt\,J(t)\phi(t)=-i\big(\alpha a^\dagger-\alpha^*a\big),\qquad \alpha\equiv\frac{i}{\sqrt{2\omega}}\int dt\,J(t)e^{i\omega t}=\frac{i\tilde J(\omega)}{\sqrt{2\omega}},
\end{equation}
so that $e^{ih[J]}=D(\alpha)$ is a displacement operator. In the vacuum,
\begin{equation}
\br{T}{G_{\rm H}}{T}=\langle h[J]^2\rangle=|\alpha|^2,\qquad \br{T_Y}{G_{\rm PJ}}{T_X}=i\,\big[h[J_Y],h[J_X]\big]=-2\,{\rm Im}\,(\alpha_Y^*\alpha_X).
\label{eq:S-mode}
\end{equation}
For a field, these expressions become sums over modes and polarizations. With $\langle\alpha,\beta\rangle\equiv\sum_{\rm modes}\alpha^*\beta$, we have $\Gamma=\frac12\|\Delta\alpha\|^2$, which is the identity $\Gamma=N_\Delta/2$ of Ref.~\cite{Matsui:2026skx}, and $\br{Y}{G_{\rm PJ}}{X}=-2\,{\rm Im}\langle\alpha_Y,\alpha_X\rangle$.

\subsection{General Gaussian influence functional}
The algebraic form of Eq.~\eqref{eq:IF} also applies to Gaussian states beyond the vacuum setting of the main theorem. Consider an environment whose influence on the matter is Gaussian, whether described by an operator field or solely by an effective action for the matter. The factor multiplying $\rho^{(0)}_{ab}$ is then the exponential of a bilinear form in $(T_a,T_b)$, with no linear terms because $\langle h\rangle=0$. When the bilinear form is expressed in terms of $\Delta$ and $\Sigma$, three requirements fix its structure:
\begin{enumerate}
\item \emph{Trace preservation}, $\rho_{aa}=\rho^{(0)}_{aa}$ for every history. Setting $T_a=T_b$ (so $\Delta=0$) must give unity, which excludes any term bilinear in $\Sigma$ alone.
\item \emph{Hermiticity}, $\rho_{ba}=\rho_{ab}^*$. Exchanging $a\leftrightarrow b$ flips $\Delta\to-\Delta$ and leaves $\Sigma$ fixed. The coefficient of the $\Delta\Sigma$ term must therefore be $i$ times a real kernel $\Rk$, and the $\Delta\Delta$ term must be real.
\item \emph{Symmetry}. Only the symmetric part $G_{\rm H}$ of the $\Delta\Delta$ kernel contributes.
\end{enumerate}
This gives the algebraic form~\eqref{eq:IF} with a real $\Rk$ and a real symmetric $G_{\rm H}$. Positivity imposes additional restrictions: these algebraic conditions alone do not ensure that the influence matrix is positive for arbitrary collections of branches. The response kernel $\Rk$ need not be retarded. For the fakeon example we adopt a tree-level projected Gaussian ansatz: the ghost contributes a time-symmetric response and no on-shell noise. This specifies the kernels used here, rather than deriving an all-orders finite-time reduced-state construction. A nonlocal mediator can be treated in the same way only after its Lorentzian response and admissible test-function space have been specified. The commutator kernel is always defined as $G_{\rm PJ}=\Rk-\Rk^T$, and the conservative kernel as $\frac12(\Rk+\Rk^T)$.

\subsection{Observables}
\emph{Visibility.} For two branches with $\Delta=T_1-T_2$, Eq.~\eqref{eq:IF} gives $|\rho_{12}|=|\rho^{(0)}_{12}|e^{-\Gamma}$ with $\Gamma=\frac12\br{\Delta}{G_{\rm H}}{\Delta}$, Eq.~\eqref{eq:Gamma}.

\emph{Conditional fringe shifts.} 
Consider two interferometers $A$ and $B$ with branch-source differences $X$ and $Y$, respectively. Since both couple linearly to the same mediator, their sources add. Fixing $A$ in branch $a_A\in\{0,1\}$, the two branches of $B$ give
\begin{equation}
\Delta=Y,\qquad
\Sigma=a_AX+\frac12Y,
\tag{S14}
\end{equation}
where a common source has been omitted. 
By Eq.~\eqref{eq:IF}, the phase of B is $(Y|R|a_AX+Y/2)$, while the noise term is independent of $a_A$. For an initially factorized state, the initial phase also cancels when comparing $a_A=0$ and $a_A=1$. Thus,
\begin{equation}
\delta\phi_{B|A}
\equiv
\arg\left(\frac{\rho_{11,10}}{\rho_{01,00}}\right)
=(Y|R|X),
\qquad
\delta\phi_{A|B}
=(X|R|Y).
\tag{S15}
\end{equation}
This is Eq.~\eqref{eq:cond}. The shift can be measured by comparing the fringes of $B$ with $A$ prepared in either branch, without requiring a superposition of $A$.

\emph{Static potential.} Consider point masses $m_A$ and $m_B$ at rest a distance $r$ apart, so that $X^{00}=m_A\delta^3(\bm x)$ and $Y^{00}=m_B\delta^3(\bm x-\bm r)$, held for a time $T$ long compared with all transients. Then
\begin{equation}
\delta\phi_{B|A}=\frac{\kappa^2}{4}m_Am_B\,T\int_{-\infty}^{\infty}d\tau\,\Rk_{00,00}(\tau,\bm r)\equiv-V_G(r)\,T .
\label{eq:S-V}
\end{equation}
The antisymmetric part of a stationary kernel is odd in $\tau$, so only the conservative part $\frac12(\Rk+\Rk^T)$ contributes. As a check, linearized GR in harmonic gauge has $\Rk_{00,00}=\frac12G_R^{(0)}$, with $G_R^{(0)}$ the retarded Green function of a massless scalar. Since $\int d\tau\,G_R^{(0)}(\tau,r)=1/4\pi r$, Eq.~\eqref{eq:S-V} gives $V_G=-\frac{\kappa^2}{4}\cdot\frac12\cdot\frac{m_Am_B}{4\pi r}=-Gm_Am_B/r$.

\emph{Radiated energy.} Let $H_0$ be the free Hamiltonian of the mediator. In the Heisenberg picture, $dH_0/dt=i[H,H_0]=\int d^3x\,J\,\partial_th$. Because the theory is linear, the expectation value of the field is given exactly by linear response, $\langle h(x)\rangle=\int d^4x'\,\Rk(x,x')J(x')$. The energy delivered to the mediator by a closed history is therefore
\begin{equation}
W=\int d^4x\,J(x)\,\partial_t\langle h(x)\rangle=\int\!\!\int d^4x\,d^4x'\,J(x)\,\partial_t\Rk(x,x')\,J(x')=\tfrac12\br{T}{\partial_t G_{\rm PJ}}{T}.
\label{eq:S-W}
\end{equation}
In the last step, we used the fact that the conservative part $\frac12(\Rk+\Rk^T)$ of a stationary kernel is even in $x-x'$. Its time derivative is then odd and vanishes under the symmetric double integral. Only the commutator part $\frac12G_{\rm PJ}$, which is Dirac's radiation-reaction kernel $\frac12(G_R-G_A)$ in a causal theory, does work. For a physically implemented source $T$, Eq.~\eqref{eq:S-W} gives its net radiative work. In condition D it is instead evaluated on $T=\Delta$, defining $W[\Delta]$. This difference-source functional includes cross terms between branch amplitudes and is generally neither the radiated energy in a single branch nor the average over branches. Measuring it requires access to those cross terms or a protocol that physically implements the difference source; single-branch recoil alone is insufficient. For a single canonical oscillator, $G_{\rm PJ}(u)=\sin(\omega u)/\omega$ and $W=\frac12|\tilde J(\omega)|^2=\omega|\alpha|^2$, the energy of $|\alpha|^2$ quanta.

\section{Spectral representation and proof of the theorem}
\label{sec:S2}

\subsection{Canonical species}
A canonical species $s$ (an ``ordinary'' species in the main text) is a free field with positive norm and positive energy, of spin $j_s\in\{0,2\}$ and mass $\mu_s$, coupled to $T^{\mu\nu}$ with unit strength. Its vacuum Wightman function, when contracted with conserved sources, is
\begin{equation}
\br{T}{W_s}{T'}=\frac{\kappa^2}{4}\int\!\frac{d^3k}{(2\pi)^32\omega_k}\,\Pi_s\!\left[\tilde T^*(\omega_k,\bm k),\tilde T'(\omega_k,\bm k)\right],\qquad \omega_k=\sqrt{\bm k^2+\mu_s^2}.
\label{eq:S-Wightman}
\end{equation}
Here $\Pi_s[A^*,B]$ denotes the on-shell polarization contraction, antilinear in $A$ and linear in $B$; its diagonal value is non-negative for a canonical species. The ordering of the conjugate follows from the stated Fourier convention. The polarization sums are given explicitly in Sec.~\ref{sec:S3}. Only positive frequencies appear, reflecting positive energy, and the weight is positive, reflecting positive norm. Since $[h,h]=-iG_{\rm PJ}$, the Wightman function decomposes as
\begin{equation}
W_s=G_{{\rm H},s}-\tfrac i2G_{{\rm PJ},s},\qquad G_{{\rm H},s}={\rm Re}\,W_s,\qquad G_{{\rm PJ},s}=-2\,{\rm Im}\,W_s .
\label{eq:S-WNE}
\end{equation}
For a single oscillator, $W=e^{-i\omega u}/2\omega$ gives $G_{\rm H}=\cos(\omega u)/2\omega$ and $G_{\rm PJ}=\sin(\omega u)/\omega$.

Spin 1 does not appear. A symmetric tensor field contains spin-2, spin-1, and two spin-0 projections. The spin-1 projector and one of the spin-0 projectors contain a factor $k_\mu$, which annihilates a conserved source. A mediator coupled to a conserved $T^{\mu\nu}$ therefore probes only the spin-2 projection and the spin-0 projection built from $\eta_{\mu\nu}-k_\mu k_\nu/k^2$, i.e., the trace.

\subsection{The representation and its scope}
The vacuum representation~\eqref{eq:spectral} assumes that the noise and commutator kernels are superpositions of the vacuum kernels of canonical species, with real, frequency-independent sector weights $\nu_s$ and $\varepsilon_s$,
\begin{equation}
G_{\rm H}=\sum_s\nu_sG_{{\rm H},s},\qquad G_{\rm PJ}=\sum_s\varepsilon_sG_{{\rm PJ},s},
\end{equation}
with sums replaced by integrals over spectral measures for continua. The representation covers a broad class of theories, including:
\begin{itemize}
\item every theory with a K\"all\'en--Lehmann representation, with weights of either sign;
\item the three free, tree-level prescriptions for quadratic gravity considered here;
\item a nonlocal theory whose specified Lorentzian prescription yields these on-shell kernels on the admitted source class.
\end{itemize}
For the explicit probe construction below, we take the massless sector to be the ordinary graviton and the additional modes to be massive. Extra massless sectors require a separate source-resolution argument. Spectrally degenerate species with the same spin and coupling are combined into one sector. The representation does \emph{not} assume that the response kernel $\Rk$ is retarded or that it has a K\"all\'en--Lehmann form; that is the content of condition C. For the stipulated vacuum in a theory with an operator field, the full Wightman function is then
\begin{equation}
W=G_{\rm H}-\tfrac i2G_{\rm PJ}=\sum_s\Big[\tfrac{\nu_s+\varepsilon_s}{2}\,W_s+\tfrac{\nu_s-\varepsilon_s}{2}\,W_s^*\Big],
\label{eq:S-Wdecomp}
\end{equation}
which shows how $\nu_s$ and $\varepsilon_s$ weight the positive-frequency part $W_s$ and the negative-frequency part $W_s^*$. In a positive-metric Gaussian operator realization, positivity of the Wightman function requires both coefficients to be non-negative, giving $\nu_s\geq|\varepsilon_s|$. Condition P, $\nu_s\geq0$, is weaker: it tests visibility rather than the full positivity of a multibranch reduced map.

\subsection{Emission functional}
For a difference source $\Delta$ we define
\begin{equation}
n_s[\Delta]=\frac{\kappa^2}{4}\int\!\frac{d^3k}{(2\pi)^32\omega_k}\Pi_s\big[\tilde\Delta^*,\tilde\Delta\big]\equiv\int_{\mu_s}^\infty\! d\omega\,\frac{dn_s}{d\omega}\geq0 .
\end{equation}
This is the mean number of quanta that would be emitted by the difference source into a canonical species, i.e., $\|\Delta\alpha\|^2$ in the mode language of Eq.~\eqref{eq:S-mode}. By Eqs.~\eqref{eq:Gamma} and \eqref{eq:S-WNE}, $\br{\Delta}{G_{{\rm H},s}}{\Delta}={\rm Re}\br{\Delta}{W_s}{\Delta}=n_s[\Delta]$. For the energy, Eq.~\eqref{eq:S-W} gives $\frac12\br{\Delta}{\partial_t G_{{\rm PJ},s}}{\Delta}$. Each mode of $W_s(x,x')$ is proportional to $e^{-i\omega_k(t-t')}$, so $\partial_tW_s=-i\omega_kW_s$ and $\partial_t G_{{\rm PJ},s}=-2\,{\rm Im}\,\partial_tW_s=2\omega_k\,{\rm Re}\,W_s$ mode by mode. The work is therefore $\omega_k$ per emitted quantum. Hence
\begin{equation}
\Gamma=\tfrac12\sum_s\nu_s\,n_s[\Delta],\qquad W[\Delta]=\sum_s\varepsilon_s\int d\omega\,\omega\,\frac{dn_s}{d\omega},\qquad \Gamma_{\rm rad}\equiv\int\frac{d\omega}{2\omega}\frac{dW}{d\omega}=\tfrac12\sum_s\varepsilon_s\,n_s[\Delta].
\label{eq:S-GW}
\end{equation}
For a single oscillator this reads $n=|\tilde J|^2/2\omega$, $W=|\tilde J|^2/2=\omega n$, and $\Gamma=n/2$.

\subsection{Lemma P: visibility positivity}
\emph{Claim:} $\Gamma[\Delta]\geq0$ for all closed histories $\Leftrightarrow$ $\nu_s\geq0$ for all $s$.

Sufficiency follows immediately from Eq.~\eqref{eq:S-GW}, since $n_s\geq0$. To establish necessity, we construct closed histories that excite essentially a single species. Exactly conserved sources can be built from superpotentials.
\begin{itemize}
\item For massive spin 0, take $T^{\mu\nu}=(\partial^\mu\partial^\nu-\eta^{\mu\nu}\Box)\Phi$ with a scalar $\Phi$.
\item For spin 2, take $T^{\mu\nu}=\partial_\alpha\partial_\beta C^{\mu\alpha\nu\beta}$ with $C$ antisymmetric in $(\mu\alpha)$ and in $(\nu\beta)$ and symmetric under pair exchange.
\end{itemize}
Both are identically conserved, $\partial_\mu T^{\mu\nu}=0$. Choose $\Phi$ (or $C$) of the form ${\rm Re}[c\,e^{ip\cdot x}]f(x/L)$, with $p$ on the mass shell of species $s$, $c$ a constant (tensor) whose projection onto spin $j_s$ is nonzero, and $f$ a smooth function of compact support. Then $\tilde T(k)$ is concentrated in a region of size $1/L$ around $k=\pm p$, and outside it decays faster than any power of $L|k\mp p|$. As $L\to\infty$, the on-shell integral $n_{s'}[T]$ of any species with $\mu_{s'}\neq\mu_s$ samples only the tails, so $n_{s'}/n_s\to0$. The polarization structure of $c$ eliminates contributions from species with the same mass but a different spin. Hence $\Gamma\to\frac12\nu_s\,n_s[T]$ with $n_s>0$, and $\Gamma\geq0$ forces $\nu_s\geq0$. For a continuous spectrum, the same argument applied to narrow mass windows shows that the spectral measure is non-negative. This construction gives exactly conserved mathematical test tensors but does not establish their realizability as differences of two positive-energy matter-and-apparatus histories. The theorem therefore assumes an admissible source class rich enough to resolve the spectral sectors; the benchmark interferometers illustrate profiles rather than establish physical source completeness. For the scalar construction, $T^\mu{}_\mu=-3\Box\Phi$ vanishes on a massless shell, so this particular probe does not resolve an additional massless scalar.

\subsection{Lemma D: decoherence--radiation identity}
\emph{Claim:} $d\Gamma/d\omega=(2\omega)^{-1}dW/d\omega$ for all closed histories $\Leftrightarrow$ $\nu_s=\varepsilon_s$ for all $s$ $\Leftrightarrow$ the Wightman function contains only positive frequencies.

By Eq.~\eqref{eq:S-GW},
\begin{equation}
\frac{d\Gamma}{d\omega}-\frac1{2\omega}\frac{dW}{d\omega}=\frac12\sum_s(\nu_s-\varepsilon_s)\frac{dn_s}{d\omega}.
\end{equation}
The localized test sources of Lemma P resolve each spectral sector, so the right-hand side vanishes for all admitted histories if and only if $\nu_s=\varepsilon_s$. Equation~\eqref{eq:S-Wdecomp} then contains no negative-frequency term $W_s^*$. In a positive-metric operator realization this is the positive-frequency spectral condition in the source-coupled sector, giving the vacuum fluctuation--dissipation relation. It does not alone prove that the full interacting Hamiltonian is bounded below. For a negative-norm realization, D still expresses a frequency-support identity even though a positive-metric ground-state interpretation is unavailable. Condition D refers to the stipulated vacuum kernels; other states are discussed in Sec.~\ref{sec:S7}.

The examples of the main text fit this statement:
\begin{itemize}
\item Negative-norm ghost: $W_{\rm ghost}=-W_{m_2}$, which has positive frequencies only, so D holds although the weight is negative.
\item Negative-energy ghost: $W_{\rm ghost}=+W_{m_2}^*$, which has negative frequencies, so D fails.
\end{itemize}

\subsection{Lemma C: causality}
\emph{Claim:} On conserved sources, condition C fixes the noncontact response to $\Rk=\theta G_{\rm PJ}$.

For a sufficiently rich class of probes, vanishing response to every strictly later source implies that the observable kernel vanishes for $y^0<x^0$. At distinct points with $y^0>x^0$, its transpose vanishes, so $\Rk(y,x)=G_{\rm PJ}(y,x)$. Conversely, the retarded kernel gives zero response to a strictly later source. Conserved probes determine these identities only modulo longitudinal kernels that vanish upon contraction.

There is also a contact ambiguity: a symmetric kernel supported at equal times contributes nothing to $G_{\rm PJ}$ and is not excluded by temporally disjoint probes. Lorentz invariance in every inertial frame restricts such an observable ambiguity to coincident points. We therefore write $\Rk=\theta G_{\rm PJ}$ modulo local contact terms. These terms do not affect the static potential at $r>0$. Likewise, the noncontact conservative kernel is $\frac12(\Rk+\Rk^T)=\frac12\sgn(t-t')G_{\rm PJ}$. With Lorentz invariance, the noncontact response has support in the causal past of the observation point.

\subsection{Static Green functions of canonical species}
For a canonical species, condition C holds, so $\Rk_s=\theta G_{{\rm PJ},s}=G_{R,s}$ is its retarded Green function. In the static limit, $\int_{-\infty}^\infty d\tau\,G_{R,s}(\tau,\bm r)$ equals the zero-frequency Green function of $-\nabla^2+\mu_s^2$, i.e., $e^{-\mu_sr}/4\pi r$, multiplied by the polarization sum evaluated on static sources. For $T^{\mu\nu}=m\,\delta^\mu_0\delta^\nu_0\delta^3(\bm x)$ we have $T\!\cdot\!T'=T^{00}T'^{00}=mm'$ and $T^\mu{}_\mu=-m$, so
\begin{equation}
\Pi_g=T\!\cdot\!T'-\tfrac12TT'=\tfrac12mm',\qquad \Pi_2=T\!\cdot\!T'-\tfrac13TT'=\tfrac23mm',\qquad \Pi_0=c_0\,TT'=c_0\,mm',
\label{eq:S-static}
\end{equation}
for the massless graviton, a massive (Fierz--Pauli) spin-2 field, and a scalar coupled to the trace. In quadratic gravity $c_0=\frac16$; in general $c_0>0$. All three are positive: static exchange of an even-spin quantum between positive masses is attractive. Multiplying by $\kappa^2/4=8\pi G$ and by $e^{-\mu_sr}/4\pi r$, each canonical species contributes to $V_G$ a term
\begin{equation}
V_s(r)=-g_s\frac{m_Am_B}{r}e^{-\mu_sr},\qquad g_s=2G\,\frac{\Pi_s}{m_Am_B}>0,
\end{equation}
that is, $g=G$ for the graviton, $g=\frac43G$ for a unit-residue massive spin-2 field, and $g=\frac13G$ for the scalar of quadratic gravity. Equation~\eqref{eq:Stelle} is the sum of these with residues $(+1,-1,+1)$.

\subsection{Proof of the theorem}
Assume P, D, and C. By Lemma C and Eq.~\eqref{eq:S-V},
\begin{equation}
V_G(r)=-\tfrac{\kappa^2}{4}m_Am_B\int_0^\infty d\tau\,(G_{\rm PJ})_{00,00}(\tau,\bm r)=-\tfrac{\kappa^2}{4}m_Am_B\sum_s\varepsilon_s\int_0^\infty d\tau\,(G_{{\rm PJ},s})_{00,00}(\tau,\bm r)=-\frac{m_Am_B}{r}\sum_s\varepsilon_s\,g_s\,e^{-\mu_sr},
\end{equation}
where the last step uses $\int_0^\infty d\tau\,G_{{\rm PJ},s}=\int d\tau\,G_{R,s}$ and the previous subsection. By Lemmas P and D, $\varepsilon_s=\nu_s\geq0$. Under the stated massless-sector assumption, the graviton contributes $\varepsilon_gg_g=\lim_{r\to\infty}r|V_G(r)|/m_Am_B=G_\infty$, Newton's constant measured at long distances. Writing $\sigma_s=\varepsilon_sg_s/G_\infty\geq0$ for the massive species gives Eq.~\eqref{eq:theorem}.

The bound then follows term by term. Since every $\sigma_se^{-\mu_sr}\geq0$, the potential obeys $|V_G(r)|\geq G_\infty m_Am_B/r$. Differentiating Eq.~\eqref{eq:theorem}, the force is
\begin{equation}
|V_G'(r)|=\frac{G_\infty m_Am_B}{r^2}\Big[1+\sum_{\mu_s>0}\sigma_s\,(1+\mu_sr)\,e^{-\mu_sr}\Big]\geq\frac{G_\infty m_Am_B}{r^2},
\end{equation}
so the force magnitude is likewise at least Newtonian. This is Eq.~\eqref{eq:bound}. The function $F(r)\equiv r|V_G(r)|/m_Am_B$ is a non-negative combination of decaying exponentials and hence non-increasing.

After subtracting local contact polynomials, in momentum space $\tilde V_G(k)=-4\pi m_Am_B\sum_s\varepsilon_sg_s/(k^2+\mu_s^2)$, so $k^2|\tilde V_G(k)|=4\pi m_Am_B\sum_s\varepsilon_sg_s\,k^2/(k^2+\mu_s^2)$. Each term increases with $k$, so $k^2|\tilde V_G|$ is non-decreasing and has a strictly positive lower bound from the massless sector; its large-$k$ limit may be infinite for a continuum with unbounded total spectral weight. A falloff faster than $1/k^2$, in particular the $1/k^4$ of a renormalizable theory, is impossible. $\square$

\emph{Remarks.} The bound is the familiar K\"all\'en--Lehmann obstruction~\cite{Kallen:1952zz,Lehmann:1954xi,Weinberg:1995mt}, applied to gravity in Ref.~\cite{Dvali:2010ue}, rather than a new static result. Our contribution is its operational decomposition into visibility positivity (P), the vacuum decoherence--radiation identity (D), and retarded response (C), which distinguish the free ghost prescriptions of Sec.~\ref{sec:S3}. These conditions diagnose the corresponding kernel properties within the stated source class, rather than establish full unitarity, interacting stability, or a complete theory of measurement. The vDVZ difference between the massless and massive spin-2 couplings ($1$ versus $\frac43$) does not affect the bound, since both static couplings are positive.

\subsection{Independence of the three conditions}
Each condition can fail while the other two hold (Sec.~\ref{sec:S3}):
\begin{itemize}
\item A negative-norm ghost has $\nu=\varepsilon=-1$ and $\Rk=\theta G_{\rm PJ}$: P fails; D and C hold.
\item A negative-energy ghost has $\nu=+1$, $\varepsilon=-1$, and $\Rk=\theta G_{\rm PJ}$: D fails; P and C hold.
\item A fakeon has $\nu=\varepsilon=0$ and $\Rk\neq\theta G_{\rm PJ}$: C fails; P and D hold.
\end{itemize}
No two of the conditions imply the third, so the trilemma cannot be reduced to a dilemma.

\section{Quadratic gravity}
\label{sec:S3}

\subsection{Propagator and species}
Quadratic gravity adds $R^2$ and $C_{\mu\nu\rho\sigma}C^{\mu\nu\rho\sigma}$ to the Einstein--Hilbert action; their coefficients set the masses $m_0$ and $m_2$~\cite{Stelle:1976gc,Stelle:1977ry}. In terms of the spin-2 and spin-0 Barnes--Rivers projectors,
\begin{equation}
P^{(2)}_{\mu\nu,\rho\sigma}=\tfrac12(\theta_{\mu\rho}\theta_{\nu\sigma}+\theta_{\mu\sigma}\theta_{\nu\rho})-\tfrac13\theta_{\mu\nu}\theta_{\rho\sigma},\qquad P^{(0)}_{\mu\nu,\rho\sigma}=\tfrac13\theta_{\mu\nu}\theta_{\rho\sigma},\qquad \theta_{\mu\nu}=\eta_{\mu\nu}-\frac{k_\mu k_\nu}{k^2},
\end{equation}
the gauge-invariant part of the Euclidean propagator is~\cite{Stelle:1977ry}
\begin{equation}
\tilde{\mathcal D}(k)=\frac{P^{(2)}-\frac12P^{(0)}}{k^2}-\frac{P^{(2)}}{k^2+m_2^2}+\frac{\frac12P^{(0)}}{k^2+m_0^2}.
\end{equation}
When contracted with conserved sources, $\theta_{\mu\nu}\to\eta_{\mu\nu}$, $T\!\cdot\!P^{(2)}\!\cdot\!T'=T\!\cdot\!T'-\frac13TT'$, and $T\!\cdot\!P^{(0)}\!\cdot\!T'=\frac13TT'$, which gives
\begin{equation}
T\!\cdot\!\tilde{\mathcal D}(k)\!\cdot\!T'=\frac{T\!\cdot\!T'-\frac12TT'}{k^2}-\frac{T\!\cdot\!T'-\frac13TT'}{k^2+m_2^2}+\frac{\frac16TT'}{k^2+m_0^2},
\label{eq:S-prop}
\end{equation}
where $T\equiv T^\mu{}_\mu$. Three species appear:
\begin{itemize}
\item the massless graviton, with polarization sum $\Pi_g=T\!\cdot\!T'-\frac12TT'$ and residue $+1$;
\item a massive spin-2 mode, with $\Pi_2=T\!\cdot\!T'-\frac13TT'$ and residue $-1$;
\item a scalar, with $\Pi_0=\frac16TT'$ and residue $+1$.
\end{itemize}
The residues are fixed by the classical action. They determine the classical retarded response, and hence $\varepsilon=(+1,-1,+1)$, in every causal quantization. In the static limit Eqs.~\eqref{eq:S-static} and \eqref{eq:S-prop} give Eq.~\eqref{eq:Stelle}, independently of how the ghost is quantized. The sum $1-\frac43+\frac13=0$ of the three static couplings expresses the $1/k^4$ falloff.

\subsection{Ghost prescriptions}
We describe each prescription by its effect on a single ghost mode of frequency $\omega$, with field $\phi_g$ and Fock operators $b,b^\dagger$.

\emph{Negative norm (indefinite metric)}~\cite{Lee:1969fy,Lee:1970iw}. The ghost is quantized with $[b,b^\dagger]=-1$ and $H=-\omega b^\dagger b$. Then $\dot b=i[H,b]=-i\omega b$, and the states $(b^\dagger)^n|0\rangle$ have positive energy $n\omega$ and norm $(-1)^nn!$. Hence $\phi_g(t)\propto be^{-i\omega t}+b^\dagger e^{i\omega t}$ and $\langle0|\phi_g(t)\phi_g(0)|0\rangle=-e^{-i\omega t}/2\omega$. This is minus the canonical Wightman function, so $\nu_2=\varepsilon_2=-1$. The coherent displacement produced by a source has characteristic function
\begin{equation}
\langle0|e^{\beta b^\dagger-\beta^*b}|0\rangle=\langle0|e^{\beta b^\dagger}e^{-\beta^*b}|0\rangle\,e^{-\frac12[\beta b^\dagger,-\beta^*b]}=e^{+|\beta|^2/2},
\end{equation}
since $[b^\dagger,b]=+1$ and $b|0\rangle=0$, instead of $e^{-|\beta|^2/2}$. Emission of a negative-norm quantum increases the overlap of the two branch states, giving rise to $\Gamma<0$.

\emph{Negative energy (positive norm)}~\cite{Woodard:2015zca,Cline:2003gs}. The ghost is quantized with $[b,b^\dagger]=+1$ and $H=-\omega b^\dagger b$. From $\dot b=i[H,b]=i\omega b$ one finds $b(t)=be^{+i\omega t}$, so $\phi_g(t)\propto be^{i\omega t}+b^\dagger e^{-i\omega t}$ and $\langle0|\phi_g(t)\phi_g(0)|0\rangle=+e^{+i\omega t}/2\omega=W_{m_2}^*$. By Eq.~\eqref{eq:S-WNE}, the Hadamard part is unchanged ($\nu_2=+1$) while the commutator changes sign ($\varepsilon_2=-1$). The classical response, and hence $\varepsilon_2$, agrees with the negative-norm case, but the fluctuations contribute additively rather than subtractively.

\emph{Fakeon (purely virtual)}~\cite{Anselmi:2018kgz,Anselmi:2018tmf}. The ghost is assigned a prescription with no on-shell cut, so it contributes neither to $G_{\rm H}$ nor to $G_{\rm PJ}$: $\nu_2=\varepsilon_2=0$. Its tree-level (classical) Green function is the time-symmetric principal-value combination $G_{\rm PV}=\frac12(G_R+G_A)$~\cite{Anselmi:2018tmf,Anselmi:2020tqo}. The matter response kernel is therefore
\begin{equation}
\Rk=G_R^{(g)}-G_{\rm PV}^{(m_2)}+G_R^{(0)},
\end{equation}
with superscripts denoting the species. Its antisymmetric part is $G_{\rm PJ}=G_{\rm PJ}^{(g)}+G_{\rm PJ}^{(0)}$, so the ghost survives only in the conservative part, and $\Rk\neq\theta G_{\rm PJ}$.

\subsection{Other completions}
The following models provide qualitative comparisons with the three prescriptions above. Their inclusion in the theorem requires a separate check of the real-mass spectral representation and of the finite-time kernel prescription.
\begin{itemize}
\item In the reversed-causality (``Merlin'') treatment of Ref.~\cite{Donoghue:2019ecz}, the unstable ghost propagates with advanced boundary conditions over its lifetime. Its contribution to $\Rk$ is therefore not retarded, which violates C.
\item Complex-conjugate Lee--Wick pairs in the CLOP prescription~\cite{Cutkosky:1969fq} have no asymptotic states and produce acausal precursors~\cite{Grinstein:2008bg}, which is analogous to failure of C. Complex poles need not admit the real-mass representation~\eqref{eq:spectral}. If complex ghosts can be produced, as argued in Ref.~\cite{Kubo:2023lpz}, the consequences for P require a separate reduced-state calculation.
\item A nonlocal model with an entire form factor can soften the static potential without adding poles, as in Ref.~\cite{Biswas:2011ar}. Pole counting alone does not determine its finite-time noise and response. If a specified Lorentzian prescription has the ordinary graviton on-shell kernels, obeys Eq.~\eqref{eq:spectral}, and is well defined on the admitted histories, P and D hold. A softened potential such as $V_G\propto-{\rm erf}(Mr/2)/r$ then requires failure of C. Particular nonlocal prescriptions exhibit microscopic acausality~\cite{Carone:2016eyp}; we do not infer this classification for every entire form factor or quantization.

\end{itemize}

\subsection{Polarization sums for slow closed histories}
For a compact, slowly moving source, the on-shell Fourier components of $T^{\mu\nu}$ can be obtained from those of the spatial stress. On the mass shell, $|\bm k|=v\omega$ with $v=(1-\mu^2/\omega^2)^{1/2}$ the speed of the emitted quantum. Conservation, $k_\mu\tilde T^{\mu\nu}=0$, then gives
\begin{equation}
\tilde T^{0i}=v\,\hat k_j\tilde T^{ij},\qquad \tilde T^{00}=v^2\,\hat k_i\hat k_j\tilde T^{ij}.
\end{equation}
In the quadrupole limit $\omega L\ll1$, the spatial stress is $\tilde T^{ij}\equiv A_{ij}=-\frac{\omega^2}{2}\Delta\tilde I_{ij}(\omega)$, where $\Delta I_{ij}=\int d^3x\,x_ix_j\,\Delta T^{00}$ is the difference of second mass moments~\cite{Matsui:2026skx}. The angular averages use $\langle\hat k_i\hat k_j\rangle=\frac13\delta_{ij}$ and $\langle\hat k_i\hat k_j\hat k_k\hat k_l\rangle=\frac1{15}(\delta_{ij}\delta_{kl}+\delta_{ik}\delta_{jl}+\delta_{il}\delta_{jk})$, which give
\begin{equation}
\langle|\hat k_i\hat k_jA_{ij}|^2\rangle=\tfrac1{15}\big(2|A|^2+|A_{kk}|^2\big),\qquad
\langle\hat k_jA_{ij}\hat k_kA^*_{ik}\rangle=\tfrac13|A|^2,\qquad
\langle(\hat k_i\hat k_jA_{ij})A^*_{kk}\rangle=\tfrac13|A_{kk}|^2,
\end{equation}
with $|A|^2=A_{ij}A^*_{ij}$. With the mostly-plus metric, $\tilde T_{\mu\nu}\tilde T^{\mu\nu*}=|\tilde T^{00}|^2-2|\tilde T^{0i}|^2+|\tilde T^{ij}|^2$ and $\tilde T^\mu{}_\mu=-\tilde T^{00}+A_{kk}$. Hence
\begin{align}
\langle\tilde T_{\mu\nu}\tilde T^{\mu\nu*}\rangle&=\tfrac{v^4}{15}\big(2|A|^2+|A_{kk}|^2\big)-\tfrac23v^2|A|^2+|A|^2,\\
\langle|\tilde T^\mu{}_\mu|^2\rangle&=\tfrac{v^4}{15}\big(2|A|^2+|A_{kk}|^2\big)-\tfrac23v^2|A_{kk}|^2+|A_{kk}|^2 .
\end{align}
Inserting them into the polarization sums and writing $\langle\Pi_s\rangle=\frac{\omega^4}{4}\mathcal W_s$ gives
\begin{align}
\mathcal W_g&=\tfrac25\big(|\Delta\tilde I|^2-\tfrac13|\Delta\tilde I_{kk}|^2\big),\nonumber\\
\mathcal W_2&=|\Delta\tilde I|^2\big(1-\tfrac23v^2+\tfrac{4}{45}v^4\big)+|\Delta\tilde I_{kk}|^2\big(\tfrac{2}{45}v^4+\tfrac29v^2-\tfrac13\big),\nonumber\\
\mathcal W_0&=|\Delta\tilde I|^2\tfrac{v^4}{45}+|\Delta\tilde I_{kk}|^2\big(\tfrac{v^4}{90}-\tfrac{v^2}{9}+\tfrac16\big),
\label{eq:S-pol}
\end{align}
where $|\Delta\tilde I|^2\equiv\Delta\tilde I_{ij}\Delta\tilde I^*_{ij}$ and $\Delta\tilde I_{kk}$ is the trace. For the graviton ($v=1$), $\mathcal W_g=\frac25|\Delta\tilde I^{\rm TF}|^2$, where TF denotes the trace-free part. This agrees with the transverse-traceless projection, since on shell $\Lambda_{ij,kl}\tilde T^{ij}\tilde T^{kl*}=\tilde T_{\mu\nu}\tilde T^{\mu\nu*}-\frac12|\tilde T|^2$~\cite{Matsui:2026skx}.

\emph{Phase space.} After angular averaging, $\int d^3k/(2\omega_k)\to2\pi\int k\,(k\,dk/\omega)=2\pi\int d\omega\,v\omega$, since $k\,dk=\omega\,d\omega$ on shell. Hence
\begin{equation}
\frac{dn_s}{d\omega}=\frac{\kappa^2}{4}\frac{2\pi}{(2\pi)^3}\,v\omega\,\langle\Pi_s\rangle=\frac{2G}{\pi}\,v\omega\,\langle\Pi_s\rangle=\frac{G}{2\pi}\,\omega^5v\,\mathcal W_s(v),
\end{equation}
and Eq.~\eqref{eq:S-GW} gives Eq.~\eqref{eq:Gamma-massive}; restoring units introduces the denominator $\hbar c^5$, with $\omega$ and the lower threshold expressed as angular frequencies. A physical mass $m_s$ has threshold $m_sc^2/\hbar$; products such as $m_s\tau$ elsewhere use natural units. For the graviton alone, Eq.~\eqref{eq:Gamma-massive} reduces to $\Gamma=\frac{G}{10\pi\hbar c^5}\int d\omega\,\omega^5|\Delta\tilde Q_{ij}|^2$, Eq.~(38) of Ref.~\cite{Matsui:2026skx}.

\emph{Sum rule.} At $v=1$,
\begin{equation}
\mathcal W_2(1)=\tfrac{19}{45}|\Delta\tilde I|^2-\tfrac1{15}|\Delta\tilde I_{kk}|^2,\qquad \mathcal W_0(1)=\tfrac1{45}|\Delta\tilde I|^2+\tfrac1{15}|\Delta\tilde I_{kk}|^2,
\end{equation}
so that
\begin{equation}
\mathcal W_g-\mathcal W_2(1)+\mathcal W_0(1)=\tfrac25|\Delta\tilde I|^2-\tfrac2{15}|\Delta\tilde I_{kk}|^2-\tfrac{18}{45}|\Delta\tilde I|^2+\tfrac2{15}|\Delta\tilde I_{kk}|^2=0,
\end{equation}
which is Eq.~\eqref{eq:sumrule}. The sum rule is the on-shell counterpart of the static identity $1-\frac43+\frac13=0$. Both express the cancellation, among the three residues, of the leading $1/k^2$ behavior of the propagator. For a rank-one source, $\Delta I_{ij}\propto\hat e_i\hat e_j$, the ratio $\mathcal W_2(1)/\mathcal W_g$ equals $4/3$, the familiar vDVZ factor.

\subsection{Benchmark histories}
The benchmark values assume that the total conserved source, including branch-dependent apparatus stresses, has a second-moment difference dominated by the stated particle trajectory. An accelerated particle alone is not a conserved source; constructing an apparatus completion and estimating its correction is a separate physical requirement. Both benchmark histories have Gaussian time dependence. The main text uses the orthogonal-arm history; the second benchmark is the collinear history of Ref.~\cite{Matsui:2026skx}. In the collinear history, branch 1 follows $\bm r_1(t)=d\,e^{-t^2/2\tau^2}\hat{\bm e}$ and branch 2 stays at the origin, so
\begin{equation}
\Delta I_{ij}=md^2e^{-t^2/\tau^2}\hat e_i\hat e_j,\qquad |\Delta\tilde I|^2=|\Delta\tilde I_{kk}|^2=\pi m^2d^4\tau^2e^{-\omega^2\tau^2/2}.
\end{equation}
In the orthogonal-arm history, $\bm r_1=d\,e^{-t^2/2\tau^2}\hat{\bm x}$ and $\bm r_2=d\,e^{-t^2/2\tau^2}\hat{\bm y}$, so
\begin{equation}
\Delta I_{ij}=md^2e^{-t^2/\tau^2}(\hat x_i\hat x_j-\hat y_i\hat y_j),\qquad |\Delta\tilde I|^2=2\pi m^2d^4\tau^2e^{-\omega^2\tau^2/2},\qquad \Delta\tilde I_{kk}=0 .
\end{equation}
Write $\nu=\omega\tau$ and let $\widehat{\mathcal W}_s$ denote $\mathcal W_s$ divided by the common spectral factor $\pi m^2d^4\tau^2e^{-\nu^2/2}$. Using $\int_0^\infty d\nu\,\nu^5e^{-\nu^2/2}=8$, Eq.~\eqref{eq:Gamma-massive} gives
\begin{equation}
\frac{\Gamma}{\Gamma_\GR}=\frac{\sum_s\nu_s\int_{\mu_s\tau}^\infty d\nu\,\nu^5v_s\,\widehat{\mathcal W}_s(v_s)\,e^{-\nu^2/2}}{8\,\widehat{\mathcal W}_g},\qquad
\Gamma_\GR=\frac{G}{4\pi}\,\pi m^2d^4\tau^2\,\frac{8}{\tau^6}\,\widehat{\mathcal W}_g=\Big\{\tfrac{8}{15},\ \tfrac85\Big\}\frac{Gm^2d^4}{\hbar c^5\tau^4},
\label{eq:S-ratio}
\end{equation}
since $\widehat{\mathcal W}_g=\frac25(1-\frac13)=\frac4{15}$ (collinear) and $\frac25\cdot2=\frac45$ (orthogonal arm). The collinear value reproduces Eq.~(49) of Ref.~\cite{Matsui:2026skx}.

\subsection{Limits}
\emph{Fast histories.} For $\tau\ll1/m_2,1/m_0$, all thresholds are negligible and $v_s\to1$ over the whole spectral weight. Define $R_s=\mathcal W_s(1)/\mathcal W_g$. For the collinear history $(R_2,R_0)=(\frac43,\frac13)$, whereas for the orthogonal-arm history $(R_2,R_0)=(\frac{19}{18},\frac1{18})$. Then $\Gamma/\Gamma_\GR\to1+\nu_2R_2+\nu_0R_0$:
\begin{itemize}
\item Negative norm: $1-R_2+R_0=0$. The limit is approached from below for the orthogonal-arm history, for every $m_0$, and from above for the collinear history when $m_0<4^{1/6}m_2\simeq1.26\,m_2$, in particular for $m_0=m_2$ (see below).
\item Negative energy: $1+R_2+R_0=\frac83$ and $\frac{19}9$.
\item Fakeon: $1+R_0=\frac43$ and $\frac{19}{18}$.
\end{itemize}
In the Einstein--Weyl limit ($m_0\gg1/\tau$) the scalar is not emitted, and the limits become $1-R_2=-\frac13$ and $-\frac1{18}$ (negative norm), $1+R_2=\frac73$ and $\frac{37}{18}$ (negative energy), and $1$ (fakeon). The difference between the two histories comes from the scalar. Since $\Pi_0\propto|\tilde T^\mu{}_\mu|^2$, it couples mainly to the trace, and $R_0$ for the trace-free history is one-sixth of its collinear value.

\emph{Approach to the fast limit.} With $s=\mu^2/\omega^2$, expanding Eq.~\eqref{eq:S-pol} gives, per unit $|\Delta\tilde I|^2$,
\begin{equation}
v\mathcal W_2=\tfrac{19}{45}+\tfrac5{18}s+O(s^2),\qquad v\mathcal W_0=\tfrac1{45}-\tfrac1{18}s+O(s^2)\qquad\text{(trace-free)},
\end{equation}
\begin{equation}
v\mathcal W_2=\tfrac{16}{45}-\tfrac19s^3+O(s^4),\qquad v\mathcal W_0=\tfrac4{45}-\tfrac1{36}s^3+O(s^4)\qquad\text{(collinear)}.
\end{equation}
For the trace-free history the negative-norm combination $-v\mathcal W_2+v\mathcal W_0$ decreases at $O(s)$ for both species, so $\Gamma$ approaches zero from below. For the collinear history the $O(s)$ and $O(s^2)$ terms vanish, and $\Gamma/\Gamma_\GR\propto\frac19(m_2\tau)^6\ln(1/m_2\tau)-\frac1{36}(m_0\tau)^6\ln(1/m_0\tau)$ at leading order, which is positive for $m_0<4^{1/6}m_2$. For a rank-one source, $\mathcal W_0(v)=\frac14\mathcal W_2(v)$ identically in $v$.

\emph{Slow histories.} For $m_2\tau\gg1$, the ghost is emitted only from the far Gaussian tail of the spectrum, near threshold. Write $x=m_2\tau$. Near threshold, $v\simeq[2(\nu-x)/x]^{1/2}$, so
\begin{equation}
\int_x^\infty d\nu\,\nu^5v\,e^{-\nu^2/2}\simeq x^5\sqrt{\tfrac2x}\,e^{-x^2/2}\int_0^\infty du\,\sqrt u\,e^{-xu}=\sqrt{\tfrac{\pi}{2}}\,x^3e^{-x^2/2}.
\end{equation}
At threshold, $\mathcal W_2(0)=|\Delta\tilde I^{\rm TF}|^2$, and therefore $\mathcal W_2(0)/\mathcal W_g=\frac52$ for every history. The ghost contribution is then
\begin{equation}
\frac{|\Gamma-\Gamma_\GR|_{\rm ghost}}{\Gamma_\GR}\simeq\frac{5\sqrt{2\pi}}{32}\,(m_2\tau)^3\,e^{-(m_2\tau)^2/2}\qquad(m_2\tau\gg1),
\label{eq:S-asym}
\end{equation}
independent of the geometry of the history. It enters with the sign of $\nu_2$. The next-order relative correction is $\frac52(m_2\tau)^{-2}+O((m_2\tau)^{-4})$, common to both benchmark geometries.

\section{Numerical results and radiated energy}
\label{sec:S4}

\subsection{Method}
The integrals~\eqref{eq:S-ratio} were evaluated by adaptive Gauss--Kronrod quadrature, with the thresholds $\nu=m_2\tau$ and $m_0\tau$ treated as breakpoints; the integrable square-root behavior at threshold needs no further treatment. Sign changes of $\Gamma$ were located by Brent's method.

\subsection{Radiated energy}
By Eq.~\eqref{eq:S-GW}, $\Gamma_{\rm rad}=\int d\omega\,(2\omega)^{-1}dW/d\omega$ is obtained from $\Gamma$ by replacing $\nu_s$ with $\varepsilon_s$. The negative-norm and negative-energy prescriptions share $\varepsilon=(1,-1,1)$ for (graviton, spin-2 ghost, scalar), so they have the same $\Gamma_{\rm rad}$, which coincides with the negative-norm value of $\Gamma$ (column 3 of Table~\ref{tab:num}), since $\nu=\varepsilon$ for a negative-norm ghost. For the fakeon, $\nu=\varepsilon=(1,0,1)$, so $\Gamma_{\rm rad}=\Gamma$ and condition D holds. For the negative-energy ghost, condition D is violated by
\begin{equation}
\Gamma-\Gamma_{\rm rad}=n_2[\Delta]\geq0,
\end{equation}
i.e., by the ghost occupation associated with the difference source. In the Einstein--Weyl limit, $\Gamma_{\rm rad}<0$ for $m_2\tau<1.675$ (collinear) and $1.574$ (orthogonal arm), and the difference-source radiative work is negative, $W[\Delta]<0$, for $m_2\tau<1.86$ and $1.77$, respectively. The difference-source work is then negative while $\Gamma\simeq2\Gamma_\GR$, because the negative-energy ghost quanta outweigh the gravitons in $W$. For $m_0=m_2$, $W[\Delta]\to0$ as $m_2\tau\to0$. It stays non-negative for the collinear history and is negative for $m_2\tau<1.74$ in the orthogonal-arm history.

\begin{table}[!htbp]
\caption{$\Gamma/\Gamma_\GR$ for the three ghost prescriptions. For the negative-norm and negative-energy ghosts, $\Gamma_{\rm rad}/\Gamma_\GR$ equals the negative-norm $\Gamma/\Gamma_\GR$ column; for the fakeon, $\Gamma_{\rm rad}=\Gamma$.}
\label{tab:num}
\begin{ruledtabular}
\begin{tabular}{lcccc}
history & $m_2\tau$ & negative norm $\Gamma/\Gamma_\GR$ & negative energy $\Gamma/\Gamma_\GR$ & fakeon $\Gamma/\Gamma_\GR$\\
\hline
\multicolumn{5}{c}{$m_0=m_2$}\\
\hline
collinear & 4 & +0.993 & +1.012 & +1.002 \\
collinear & 3 & +0.884 & +1.193 & +1.039 \\
collinear & 2 & +0.430 & +1.950 & +1.190 \\
collinear & 1.5 & +0.169 & +2.385 & +1.277 \\
collinear & 1 & +0.030 & +2.616 & +1.323 \\
collinear & 0.5 & +0.001 & +2.665 & +1.333 \\
collinear & 0.1 & +0.000 & +2.667 & +1.333 \\
\hline
orthogonal arm & 4 & +0.990 & +1.010 & +1.000 \\
orthogonal arm & 3 & +0.848 & +1.153 & +1.000 \\
orthogonal arm & 2 & +0.273 & +1.738 & +1.006 \\
orthogonal arm & 1.5 & -0.020 & +2.049 & +1.015 \\
orthogonal arm & 1 & -0.110 & +2.171 & +1.030 \\
orthogonal arm & 0.5 & -0.046 & +2.142 & +1.048 \\
orthogonal arm & 0.1 & -0.002 & +2.112 & +1.055 \\
\hline
\multicolumn{5}{c}{Einstein--Weyl limit $m_0\to\infty$}\\
\hline
collinear & 4 & +0.990 & +1.010 & +1.000 \\
collinear & 3 & +0.845 & +1.155 & +1.000 \\
collinear & 2 & +0.240 & +1.760 & +1.000 \\
collinear & 1.5 & -0.108 & +2.108 & +1.000 \\
collinear & 1 & -0.293 & +2.293 & +1.000 \\
collinear & 0.5 & -0.332 & +2.332 & +1.000 \\
collinear & 0.1 & -0.333 & +2.333 & +1.000 \\
\hline
orthogonal arm & 4 & +0.990 & +1.010 & +1.000 \\
orthogonal arm & 3 & +0.847 & +1.153 & +1.000 \\
orthogonal arm & 2 & +0.267 & +1.733 & +1.000 \\
orthogonal arm & 1.5 & -0.034 & +2.034 & +1.000 \\
orthogonal arm & 1 & -0.141 & +2.141 & +1.000 \\
orthogonal arm & 0.5 & -0.094 & +2.094 & +1.000 \\
orthogonal arm & 0.1 & -0.057 & +2.057 & +1.000 \\
\end{tabular}
\end{ruledtabular}
\end{table}

\subsection{Summary of the numerical results}
\emph{Negative-norm thresholds.} For the orthogonal-arm history, $\Gamma<0$ for $m_2\tau<1.451$, $1.546$, and $1.574$ when $m_0/m_2=0.1$, $1$, and $\infty$, respectively. The corresponding minima are $\Gamma/\Gamma_\GR=-0.086$, $-0.111$, and $-0.141$, reached near $m_2\tau\simeq1$. For the collinear history, the larger scalar coupling ($R_0=\frac13$) keeps $\Gamma\geq0$ for all $m_2\tau$ when $m_0\lesssim1.26\,m_2$. For larger $m_0$, $\Gamma<0$ for $m_2\tau$ below $1.341$, $1.646$, and $1.675$, with minima $-0.017$, $-0.126$, and $-0.255$, when $m_0/m_2=1.5$, $2$, and $3$. For $m_0\geq10\,m_2$, $\Gamma<0$ for $m_2\tau<1.675$, with minimum $-0.332$ at $m_0=10\,m_2$, which tends to $-\frac13$ as $m_0\to\infty$.

\emph{Negative energy.} For $m_0\to\infty$, $\Gamma_{\rm rad}<0$ for $m_2\tau<1.675$ (collinear) and $1.574$ (orthogonal arm), and $W<0$ for $m_2\tau<1.86$ and $1.77$, while $\Gamma\simeq2\Gamma_\GR$. For $m_0=m_2$, $\Gamma_{\rm rad}\to0$ while $\Gamma\to\frac83\Gamma_\GR$ or $\frac{19}9\Gamma_\GR$. In both cases the matter decoheres more strongly than in GR while the difference-source radiative work is small or negative.

\emph{Fakeon.} $\Gamma=\Gamma_{\rm rad}$ equals the sum of the GR and scalar contributions for all $m_2\tau$. The ghost leaves no trace in single-interferometer decoherence or in the radiated energy.

\section{Reciprocal wake of a purely virtual ghost}
\label{sec:S5}

\subsection{Configuration}
Consider two identical, spatially compact source profiles, with $A$ delayed by $\delta\tau$ relative to $B$. The ghost-only identities below require only temporally ordered supports, with every point of $A$ later than every point of $B$. Eliminating massless GR from the full shifts additionally requires strict timelike separation of the spacetime supports. In four-dimensional flat spacetime, the massless retarded Green function is supported on the light cone, so it gives no coupling under this stronger condition. For spatially extended sources, a sufficient condition is that the temporal gap exceed the largest light-travel time between points in the two supports.

The Gaussian benchmark profiles are not compactly supported and never have strictly disjoint supports. At $\delta\geq4$, their overlap has a small Gaussian factor, but this does not by itself bound every conditional response. Figure~\ref{fig:wake} displays the on-shell commutator-based massive wake component, interpreted using the separated-support formulas; it does not include all conservative or near-field terms associated with the overlapping tails. Those terms require specified spatial profiles and a direct convolution with the full response.

A massive scalar Green function has a retarded part
\begin{equation}
G_R^{(m)}(t,\bm r)=\frac{\theta(t)}{4\pi r}\delta(t-r)-\frac{m\,\theta(t-r)}{4\pi}\frac{J_1\big(m\sqrt{t^2-r^2}\big)}{\sqrt{t^2-r^2}},
\end{equation}
whose second term is the timelike wake. At fixed $r$ it oscillates with frequency $m$ and decays as $t^{-3/2}$.

\subsection{Conditional shifts}
For a canonical massive spin-2 mode with unit residue, define the wake coupling
\begin{equation}
K\equiv\br{X}{G_{\rm PJ}^{(m_2)}}{Y}.
\end{equation}
In units of $4\Gamma_\GR$ it reads
\begin{equation}
\frac{K}{4\Gamma_\GR}=\frac{\int_{m_2\tau}^\infty d\nu\,\nu^5v\,\widehat{\mathcal W}_2(v)\,e^{-\nu^2/2}\sin(\nu\delta)}{8\widehat{\mathcal W}_g}.
\label{eq:S-wake}
\end{equation}
To derive Eq.~\eqref{eq:S-wake}, note that for identical histories with $A$ delayed, $J_X(t)=J_Y(t-\delta\tau)$, so that $\alpha_X=e^{i\omega\delta\tau}\alpha_Y$ mode by mode. Equation~\eqref{eq:S-mode} then gives
\begin{equation}
\br{X}{G_{\rm PJ}}{Y}=-2\,{\rm Im}\langle\alpha_X,\alpha_Y\rangle=2\sum_{\rm modes}\sin(\omega\delta\tau)\,|\alpha_Y|^2 ,
\end{equation}
while $4\Gamma_\GR=2\|\alpha_Y\|_g^2$. Relative to the integrand of $\Gamma$ in Eq.~\eqref{eq:S-ratio}, each frequency is therefore weighted by $\sin(\nu\delta)$, and the massive mode enters with its threshold and velocity factor. For the massless graviton the same steps, together with
\begin{equation}
\int_0^\infty d\nu\,\nu^5e^{-\nu^2/2}\sin(\nu\delta)=\sqrt{\tfrac{\pi}{2}}\,He_5(\delta)e^{-\delta^2/2},\qquad He_5(\delta)=\delta^5-10\delta^3+15\delta ,
\end{equation}
give $\br{X}{G_{\rm PJ}}{Y}_\GR/4\Gamma_\GR=\sqrt{\pi/2}\,He_5(\delta)e^{-\delta^2/2}/8$. This quantity equals $0.023$ at $\delta=4$ and decreases monotonically for $\delta\geq4$. This is the commutator bilinear from the Gaussian tails and is not included in Fig.~\ref{fig:wake}. It is not an upper bound on each individual conditional shift: the commutator fixes the difference of the two retarded responses, while overlap-dependent conservative terms also contribute.

For temporally ordered supports with $X$ entirely later than $Y$, $\br{X}{G_R}{Y}=\br{X}{G_{\rm PJ}}{Y}$ and $\br{Y}{G_R}{X}=0$ for any causal kernel. The ghost contribution to the response kernel then yields the following conditional shifts.
\begin{itemize}
\item \emph{Causal ghost (negative norm or negative energy).} Here $\Rk^{(2)}=-G_R^{(m_2)}$, so
\begin{equation}
\delta\phi_{A|B}=\br{X}{\Rk^{(2)}}{Y}=-K,\qquad \delta\phi_{B|A}=\br{Y}{\Rk^{(2)}}{X}=0 .
\end{equation}
Only the later interferometer can respond, so any nonzero response is maximally nonreciprocal.
\item \emph{Fakeon.} Here $\Rk^{(2)}=-\frac12(G_R+G_A)^{(m_2)}$. Using $G_A(x,y)=G_R(y,x)$,
\begin{align}
\delta\phi_{A|B}&=-\tfrac12\br{X}{G_R+G_A}{Y}=-\tfrac12\br{X}{G_R}{Y}=-\tfrac12K,\nonumber\\
\delta\phi_{B|A}&=-\tfrac12\br{Y}{G_R+G_A}{X}=-\tfrac12\br{X}{G_R}{Y}=-\tfrac12K .
\end{align}
Comparing with the causal case gives Eq.~\eqref{eq:reciprocal}. The causal ghost response is split evenly between the past and the future, as in Wheeler--Feynman exchange~\cite{Wheeler:1945ps,Anselmi:2020tqo}.
\end{itemize}
\emph{Scalar wake.} The scalar is a causal species with $\varepsilon_0=\nu_0=+1$ in all three prescriptions. For temporally ordered supports it adds $K_0\equiv\br{X}{G_{\rm PJ}^{(0)}}{Y}$, given by the analogous scalar spectral bilinear, to $\delta\phi_{A|B}$ only. For a rank-one source $\mathcal W_0=\frac14\mathcal W_2$ (Sec.~\ref{sec:S3}), so a collinear history with $m_0=m_2$ has $K_0=K/4$. A fakeon then gives $\delta\phi_{A|B}=-\frac14K$ and $\delta\phi_{B|A}=-\frac12K$, i.e., a ratio of $2$. The scalar commutator wake is small in absolute magnitude for the trace-free benchmark histories, where $\mathcal W_0\propto v^4$ near threshold, and for $m_0\gtrsim2m_2$, where it is suppressed by roughly $e^{-(m_0^2-m_2^2)\tau^2/2}$. Its Gaussian benchmark value follows from Eq.~\eqref{eq:S-wake} with $\widehat{\mathcal W}_0$ and threshold $m_0\tau$. For the Gaussian commutator components at $m_2\tau=2$ and $4\leq\delta\leq16$, the maximum of $|K_0|$ is $0.4\%$ of that of $|K|$ for a trace-free history with $m_0=m_2$, and $1\%$ for a collinear history with $m_0=2m_2$.

For temporally ordered, disjoint supports with $A$ later than $B$, the advanced shift $\delta\phi_{B|A}$ vanishes for a causal mediator and equals $-\frac12K$ for the adopted fakeon prescription, whenever the ghost wake is nonzero. The Gaussian plots isolate this wake component and do not establish an exact vanishing of overlap contributions. With strictly timelike-separated supports, where the massless contribution vanishes, and negligible scalar wake, the ratio $\delta\phi_{B|A}/\delta\phi_{A|B}$ is $0$ for a causal ghost and approximately $1$ for a fakeon, provided the denominator is nonzero. Including the scalar but neglecting the massless contribution, the ratio is $K/(K-2K_0)$. Even a small scalar wake can change it substantially near a node of $K$.

\subsection{Size and long-delay behavior}
The wake is nearly the same for the two benchmark histories because their ratios $\mathcal W_2/\mathcal W_g=\frac52-\frac53v^2+cv^4$, with $c=\frac29$ (orthogonal arm) or $\frac12$ (collinear), differ only in the $v^4$ term. For $m_2\tau=2$, the first extremum of $|K|$ for $\delta\geq4$ is $0.149\times4\Gamma_\GR$ at $\delta\simeq4.3$ for either history, so the inferred fakeon wake component is $0.074\times4\Gamma_\GR$ (Fig.~\ref{fig:wake}). For $m_2\tau=1$ the wake is weaker and slower: $|K|$ falls from $0.045\times4\Gamma_\GR$ ($0.047$ for the collinear history) at $\delta=4$, and its first extremum beyond $\delta=4$ is $0.010\times4\Gamma_\GR$ at $\delta\simeq8.1$. For long delays, the integral~\eqref{eq:S-wake} is dominated by the threshold $\nu\to m_2\tau$, where $v\propto\sqrt{\nu-m_2\tau}$. Writing $x=m_2\tau$, the endpoint contribution gives
\begin{equation}
\frac{K}{4\Gamma_\GR}\simeq\frac{\sqrt{2\pi}}{2}\,\frac{5}{16}\,x^{9/2}e^{-x^2/2}\,\delta^{-3/2}\sin(x\delta+3\pi/4)\qquad(\delta\gg1),
\end{equation}
whose absolute envelope scales as $\delta^{-3/2}$. For $x=2$, the checked long-delay extrema at $\delta\gtrsim20$ agree with this asymptotic formula to within $0.5\%$. Operationally, the fringe of the earlier interferometer $B$ depends on whether the later interferometer $A$ is prepared in branch $0$ or $1$. For temporally ordered, disjoint physical supports, this would be an advanced response in the adopted kernel prescription. The Gaussian commutator calculation alone does not determine the full fringe shift. For these Gaussian commutator wake components, the amplitude contains $e^{-(m_2\tau)^2/2}$ and the long-delay envelope decays as a power inherited from the Bessel tail of $G_A^{(m_2)}$ in the tree-level treatment used here. Other switching profiles need not share this exponential suppression. For these Gaussian benchmark profiles, the wake component is exponentially suppressed when $m_2\tau\gg1$~\cite{Anselmi:2018tmf}. Whether a finite ghost width generated at loop level cuts off the power-law tail is beyond the scope of this work.

\section{Physical scales}
\label{sec:S6}

For a single Yukawa correction with $|\alpha|=1$, Ref.~\cite{Lee:2020zjt} reports $\lambda<38.6\,\mu$m at $95\%$ confidence. Using this as an indicative range gives $m_2c^2\sim\hbar c/\lambda\simeq5.1$\,meV and $\hbar/(m_2c^2)\sim1.3\times10^{-13}$\,s. This is a benchmark scale rather than a fitted lower bound on $m_2$: Eq.~\eqref{eq:Stelle} has a spin-2 strength $-4/3$ and an additional scalar Yukawa term, so a quantitative constraint requires fitting both terms. At this benchmark, $m_2c^2\tau/\hbar\lesssim2$ corresponds to $\tau\lesssim0.26$\,ps, and the negative-norm sign change occurs near $\tau\simeq0.2$\,ps.

For the center-of-mass motion of matter-wave interferometers, $\tau$ ranges from milliseconds to $10^2$\,s~\cite{Matsui:2026skx}. Even for $\tau=1$\,ns, $m_2\tau\simeq7.8\times10^3$, and Eq.~\eqref{eq:S-asym} gives a suppression of order $(m_2\tau)^3e^{-(m_2\tau)^2/2}\sim10^{-1.3\times10^{7}}$. In addition, $\Gamma_\GR$ itself is of order $10^{-61}$ for the parameters of proposed gravitational-entanglement experiments~\cite{Matsui:2026skx}. The dynamical fingerprints are therefore far beyond the reach of these experiments.

The static modification is appreciable at $r\sim\hbar/(m_2c)$, about $40\,\mu$m for this benchmark and shorter for larger masses. These separations are comparable to those in proposed gravitational-entanglement experiments~\cite{Bose:2017nin,vanManen:2026akg} and to those probed by torsion-balance tests~\cite{Lee:2020zjt}, where the bound~\eqref{eq:bound} can be tested directly.

\section{Scope of the assumptions}
\label{sec:S7}

\emph{Linear coupling.} The operator influence functional~\eqref{eq:IF} is exact for a free or generalized-free Gaussian mediator with linear coupling; an interacting field need not remain Gaussian simply because its coupling to the source is linear. For gravity this is the tree-level, $O(G)$ description. Corrections beyond this order are of two kinds. Classical graviton self-interactions produce non-Gaussian terms suppressed by the post-Minkowskian parameter $Gm/(c^2L)$ of the source, with $L$ its size. In the low-energy Einstein regime, quantum loops carry factors of $(\omega t_{\rm P})^2$, with $t_{\rm P}$ the Planck time. Near the extra poles of quadratic gravity, perturbative control also depends on the curvature-squared couplings and any widths; the low-energy estimate alone does not justify the free approximation in that regime. The theorem itself applies to any mediator whose two-point kernels admit the representation~\eqref{eq:spectral}, including loop-corrected spectral functions, as long as higher cumulants of the influence functional are negligible.

\emph{Universality.} The theorem assumes that the mediator couples to the conserved, universal stress tensor $T^{\mu\nu}$. A vector boson coupled to a conserved current such as baryon number mediates a repulsive force between like charges with positive norm and energy. Such a force can weaken the total interaction at short distances without contradicting the theorem, but it is composition dependent and can be separated from gravity by equivalence-principle tests.

\emph{Ordinary mass shells and matter vertices.} The representation~\eqref{eq:spectral} assumes that the noise and commutator kernels are superpositions of the kernels of free particles with definite mass and spin, coupled through the local vertex~\eqref{eq:S-Hint}. The positive Yukawa form~\eqref{eq:theorem} then rests on this representation together with universal coupling and linear response. Momentum-dependent vertices on the matter side change the assumed local source coupling and may smear the effective source while preserving linearity. Nonlinear screening instead makes the response depend on the source amplitude. Neither effect is described by the fixed local point-source exchange used to derive Eq.~\eqref{eq:theorem}. A weakening due to either effect therefore need not imply failure of P, D, or C. Nonlocal mediator form factors are covered only if a specified Lorentzian prescription defines the kernels on the admitted histories and satisfies Eq.~\eqref{eq:spectral}. If the on-shell noise and commutator remain ordinary while the static response is softened, the noncontact conservative response is incompatible with C. Entire form factors that do not define suitable distributions on the chosen source space require a separate treatment. The theorem constrains the linear, universal exchange between separated sources within these assumptions.

\emph{Other states.} The main analysis uses Lorentz-invariant, stationary Gaussian Fock reference kernels, or projected kernels for a fakeon; the term ``vacuum'' does not imply a ground state. The Gaussian influence-functional algebra also applies to other states. For an ordinary canonical species in a thermal state at inverse temperature $\beta$, the noise weight becomes frequency dependent, $\nu_s(\omega)=\coth(\beta\omega/2)$, while $\varepsilon_s=1$, giving $d\Gamma/d\omega=\coth(\beta\omega/2)(2\omega)^{-1}dW[\Delta]/d\omega$. Squeezed Gaussian states are generally nonstationary~\cite{Matsui:2026skx}, so $G_{\rm H}$ is not time-translation invariant and cannot be represented by the constant sector weights of Eq.~\eqref{eq:spectral}. Such state-dependent changes to D do not by themselves diagnose a ghost or an instability; the mediator state must be controlled before applying the vacuum classification.

\emph{Closed histories.} All observables are built from closed histories of finite duration. For conserved smooth compact histories the kernel formulas can be evaluated without an $S$ matrix. Gaussian profiles are convenient rapidly decaying benchmarks, not strictly finite-duration sources. Operational interpretation also requires the physical source realization and, for projected theories, the finite-time kernel prescription specified above. The theorem constrains the mediator through its response, commutator, and noise kernels, which are exactly the quantities that tabletop tests of quantum gravity access.

\end{document}